\documentclass[12pt,a4paper]{article}
\usepackage[german, english]{babel}
\usepackage{a4wide}
\usepackage{amsmath}
\usepackage{amssymb}
\usepackage{ifthen}
\usepackage{epsfig}
\newcounter{fig}   \newcommand{\lbfig}[1]{\refstepcounter{fig}
\label{#1} } 
\newcommand{\vphi}{\varphi}
\newcommand{\Tr}{{\rm Tr}}

\def\vecphi{{\pmb{\varphi}}}
\def\vecsigma{{\pmb{\sigma}}}

\def\vectheta{{\pmb{\theta}}}

\begin{document}

\begin{titlepage}
\vspace*{2cm}

\begin{center}
{\bf\large Multimonopoles and closed vortices in $SU(2)$ Yang-Mills-Higgs theory}
\vspace{2.0cm}

{\sc\large Ya. Shnir}\\[12pt]
{\it  Institute of Physics, University of Oldenburg}\\
{\it D-26111, Oldenburg, Germany}

\end{center}

\date{~}

\bigskip

\begin{abstract}
We review classical monopole solutions of the $SU(2)$ Yang-Mills-Higgs 
theory. The first part is a pedagogical introduction into 
to the basic features of the celebrated 't~Hooft - Polyakov 
monopole. In the second part we describe new classes of static axially symmetric solutions
which generalise 't~Hooft - Polyakov monopole. These configurations are either
deformations of the topologically trivial sector or the sectors with
different topological charges. In both situations we construct the    
solutions representing the chains of monopoles and antimonopoles in static equilibrium. 
The solutions of another type are closed vortices which are centred around the symmetry 
axis and form different bound systems. Configurations of the third type are monopoles 
bounded with vortices. We suggest classification of these solutions which is 
related with $2d$ Poincare index.
\end{abstract}
\bigskip


\end{titlepage}



\bigskip

\section*{Introduction}
There are two completely distinct periods in the long history of the monopole
problem. Soon after the pioneering work by Dirac \cite{Dirac31}, the interest
in this field of research faded away although some enthusiasts still
worked on it 
\footnote{For the complete discussion of `stone age history' of the 
monopole problem, the properties of the 
Dirac monopole and exhaustive bibliography before 1974 
see  \cite {Str-Tom,Giac-bibl,Gold-Trower}. Note that in \cite{Gold-Trower} 
the genesis of the monopole problem was traced up to the notes by
Petrus Pelegrinius, written at the Crusades in 1269! 
We will not go into this fascinating story.}. 
Actually, for more then 40 years the monopole problem was
considered to be a rather esoteric, fairly beyond mainstream of theoretical physics of
that time. Such an attitude was caused in part by negative results of all the
experiments searching for monopoles, and it was reinforced by rather cumbersome
character of the theoretical constructions which were connected with the
corresponding generalisation of quantum electrodynamics. As a matter of fact
this problem still remains unsolved and, within an Abelian theory, a monopole
looks like a stranger. Indeed, a magnetic monopole could be introduced
into the Abelian electrodynamics on the classical level if a vector
potential is not defined globally or if there are singular objects in the
theory (for related discussion see e.g. \cite{Str-Tom}). 
However, even in that case the quantum theory of the monopole is full
of contradictions  one can hardly avoid in the framework of an Abelian
model. 

The situation changes drastically if we take into account that the Abelian
electrodynamics is not a lonely theory but a part of an unified model, 
i.e., the generator of the electromagnetic $U(1)$ subgroup has to be 
embedded into a non-Abelian gauge group of higher rank.

The modern era of the monopole theory started in 1974 when 't~Hooft and
Polyakov independently discovered monopole solutions of the $SO(3)$
Georgi--Glashow model \cite{Hooft74,Polyakov74}. Preskill pointed out that 
the essence of this
breakthrough is that while a Dirac monopole {\it could be} incorporated in an
Abelian theory, some non-Abelian models, like that of Georgi and Glashow, {\it
inevitably contain} monopole-like solutions \cite{Preskill84}. 

For many years, starting from the pioneering paper by Dirac, the most serious
argument to support the monopole concept, apart from his emotional belief that 
``one would be surprised if Nature had made no use of it'' \cite{Dirac31}, 
was the possible explanation of the quantisation of the electric charge. 
But as time went on and the idea of 
grand unification emerged, it seemed that the latter argument has lost some power.
Indeed, the modern point of view is that the operator of electric charge is
the generator of a $U(1)$ group. The charge quantisation condition arises in
models of unification if the electromagnetic subgroup is embedded into a
semi-simple non-Abelian gauge group of higher rank. In this case the electric
charge generator forms nontrivial commutation relations with all other
generators of the gauge group. Therefore, the electric charge quantisation
today is considered as an argument in support of the unification approach.

However it turns out that both the `old' and `new' explanations of the electric
charge quantisation are just two sides of the same problem, because it was
realised that almost any theory of unification with an electromagnetic $U(1)$ subgroup
embedded into a gauge group, which is spontaneously broken by the Higgs mechanism, possesses
monopole-like solutions. 
An example of such a model is the $SU(2)$ Yang-Mills-Higgs (YMH) theory which we
shall consider below.

\section{'t~Hooft - Polyakov monopole}
\subsection{$SU(2)$ Yang-Mills-Higgs model}
We consider non-Abelian classical Lagrangian which describes
coupled gauge and Higgs fields:
\begin{eqnarray}           \label{LagrGG}
L &=& -\frac{1}{2} \Tr F_{\mu\nu}F^{\mu\nu} + \Tr D_\mu \phi D^\mu \phi 
- V(\phi) \nonumber\\ 
&=& -\frac{1}{4} F_{\mu \nu}^a F^{a \mu \nu} 
+ \frac{1}{2} (D^{\mu} \phi ^a) (D_{\mu} \phi ^a )  - V(\phi) \ .
\end{eqnarray}
Here, $F_{\mu\nu} = F_{\mu \nu}^a T^a$, $\phi = \phi^a T^a$ and we use
standard normalisation of the generators of the gauge group: $\Tr (T^a T^b) =
\frac{1}{2}\delta_{ab}$, $a,b = 1,2,3$, which 
satisfy the Lie algebra 
\begin{equation}
[T^a,T^b]=i\varepsilon_{abc}T^c \ .
\end{equation}

In what follows we will choose the simplest non-trivial case of 
$SU(2)$ group \footnote{Note that the simply connected covering $SU(2)$ 
group is locally isomorphic to the $SO(3)$. In the monopole theory
the global difference between these two groups is very important because the
topological properties of the corresponding group spaces are different. For the sake 
of brevity we do not discuss this point and refer the reader to \cite{Preskill84}.},
thus the generators  could be taken in the
fundamental representation, $T^a=\frac{\sigma^a}{2}$ or in the adjoint
representation, $(T^a)_{bc}=-i\varepsilon_{abc}$ respectively. 
The covariant derivative is defined as
\begin{equation}                   \label{Eq:covar-der}
D_{\mu} = \partial _{\mu} + ie A_\mu \ ,
\end{equation}
which yields
\begin{equation}                   \label{D-der}
D_\mu \phi = \partial _{\mu}\phi + ie [A_\mu,\phi],\quad {\rm or}\quad
D_{\mu} \phi ^a = \partial _{\mu}\phi ^a 
- e \varepsilon _{abc} A_{\mu}^b  \phi ^c \ ,
\end{equation}
and the potential of the scalar fields is taken to be
\begin{equation}                    \label{potV}
V(\phi) = \frac{\lambda}{4} (\phi ^a \phi ^a - a^2)^2 \ ,
\end{equation}
where $e$ and $\lambda$ are gauge and scalar coupling constants respectively. 
The field strength tensor is  
\begin{equation}                      \label{F}
F_{\mu \nu}^a = \partial _{\mu}A_{\nu}^a 
- \partial_{\nu}A_{\mu}^a - e\varepsilon_{abc} A_{\mu}^b A_{\nu}^c \ ,
\end{equation}
or in matrix form
\begin{equation}
F_{\mu \nu} = \partial _{\mu}A_{\nu} 
- \partial _{\nu}A_{\mu} + ie[A_\mu ,A_\nu] 
\equiv \frac{1}{ie}[D_{\mu}, D_{\nu}] \ .
\end{equation}

The field equations corresponding to the Lagrangian (\ref{LagrGG}) 
\begin{equation}                      \label{Yang-Mills}
D_{\nu} F^{a \mu \nu} =  - e \varepsilon_{abc} \phi ^b D^{\mu} \phi ^c \ ;
\qquad 
D_{\mu}D^{\mu} \phi ^a = - \lambda \phi ^a (\phi ^b \phi ^b - a^2)\ .
\end{equation}
The symmetric stress-energy tensor $T_{\mu \nu}$ which follows from the
Lagrangian (\ref{LagrGG}) and the field equations (\ref{Yang-Mills}) is
\begin{eqnarray}                  \label{enegry-stress}
T_{\mu \nu} &=& 
 F_{\mu \rho}^a F^{a \rho}_\nu + (D_{\mu} \phi ^a)
(D_{\nu} \phi ^a ) - g_{\mu \nu}L \nonumber\\
&=& F_{\mu\alpha}^a F^{\nu\alpha a} + D_{\mu}\phi^aD_{\nu}\phi^a
-\frac{1}{2} g_{\mu\nu} D_\alpha\phi^a D^\alpha\phi^a 
-\frac{1}{4} g_{\mu\nu}F_{\alpha\beta}^a F^{\alpha\beta a}  \nonumber\\
&-& g_{\mu\nu}\frac{\lambda}{4}\left(\phi^2 - a^2\right) \ ,
\end{eqnarray}
and is conserved by virtue of the field equations:
$
\partial_\mu T^{\mu \nu} = 0 
$.
From (\ref{enegry-stress}) we can easily obtain the static Hamiltonian
\begin{align}                                       \label{En}
E &= \int d^3 x\, T_{00} \!=\! \int d^3 x \left[
\frac{1}{4} F_{\mu \nu}^a F^{\mu \nu a} 
\!+\! \frac{1}{2} (D_\mu \phi ^a)(D^\mu \phi ^a )
\!+\! \frac{\lambda}{4} (\phi ^a \phi ^a \!-\! a^2)^2\right]\nonumber\\
&=\int d^3 x \frac{1}{2}\left[ E_n^a E_n^a + 
B_n^a B_n^a + (D_{n} \phi ^a)(D_{n} \phi ^a )\right] + V(\phi) \ , 
\end{align}
where 
\begin{equation}
E_n^a \equiv F_{0n}^a, \qquad \text{and} \qquad
B_n^a \equiv \frac{1}{2} \varepsilon_{nmk} F_{mk}^a
\end{equation}
are `colour' electric and magnetic fields.  We see that the energy is minimal
if the following conditions are satisfied
\begin{equation}                       \label{higgs-vacuum}
\phi ^a \phi ^a = a^2;\qquad F_{m n}^a = 0;\qquad D_n \phi ^a = 0 \ .
\end{equation}
These conditions define the vacuum. 

The perturbative spectrum of the model can be found from analysing small
fluctuations around the vacuum state. Let us suppose that the system under
consideration is static, $E_n^a = 0$. Then the energy of the vacuum is equal
to zero. We consider now a fluctuation $\chi$ of the scalar field
$\phi$ around the trivial vacuum $|\phi| = a$ where only the third isotopic
component of the Higgs field is non-vanishing:
\begin{equation}              \label{phi-fluct}
\phi = (0, 0, a+\chi) \ .
\end{equation}
Substitution of the expansion (\ref{phi-fluct}) into the Lagrangian
(\ref{LagrGG}) yields, up to terms of the second order
\begin{equation}  
(D_{n} \phi ^a)(D_{n} \phi ^a ) \approx 
(\partial_{n} \chi ^a)(\partial_{n} \chi^a ) + e^2a^2 
\left[\left(A_n^1\right)^2 + \left(A_n^2\right)^2 \right] \ ,  
\end{equation}
and
\begin{equation}
V(\phi) \approx \frac{\lambda}{2} a^2 \chi^2 \ .
\end{equation}
Thus, the vacuum average of the scalar field is non-vanishing and the model
describes spontaneous symmetry breaking. Further analysis 
shows that the perturbative spectrum consists of a
massless photon $A_\mu^3$ corresponding to the unbroken $U(1)$ electromagnetic
subgroup, massive vector fields $A^{\pm}_\mu = (1/\sqrt 2)(A^1_\mu \pm A^2_\mu)$ with mass
$m_v = ea$, and neutral scalars having a mass $m_s=a\sqrt{2\lambda}$.
 
Note 
that the electric charge of the massive vector bosons $A^{\pm}$ is given by
the unbroken $U(1)$ subgroup. In general, this is a subgroup $H$ of the gauge
group $G$, the action of which leaves the Higgs vacuum invariant. Obviously,
that is a little group of the rotation in isospace about the direction given
by the vector $\phi^a$. The generator of it, $(\phi^a T^a)/a$, must be
identified with the operator of electric charge $Q$. Thus, the expression
for the covariant derivative (\ref{D-der}) can be written in the form
\begin{equation}
D_\mu = \partial_{\mu} + ie A_{\mu}^a T^a 
= \partial_{\mu} + iQ A_{\mu}^{\rm em}
\end{equation}
that allows to define the `electromagnetic projection' of the gauge
potential 
\begin{equation}                           \label{A-em}
A_\mu^{\rm em} = \frac{1}{a}A_{\mu}^a \phi^a,  \qquad 
Q = e\frac{1}{a} \phi^a T^a.
\end{equation}
Taking into account the definition of the generators $T^a$ of the gauge group,
we can easily see that the minimal allowed eigenvalues of the electric charge 
operator now are $q = \pm e/2$.

\subsection{Topological classification of the solutions}
 
The spectrum of possible solutions of the Yang-Mills-Higgs model is much richer
than one would naively expect. There are stable soliton-like static solutions
of the complicated system of field equations (\ref{Yang-Mills}) having a
finite energy density on the spatial asymptotic. An adequate description of
these objects needs the extensive use of topological methods.

Indeed, the very definition (\ref{higgs-vacuum}) forces the classical vacuum
of the Yang-Mills-Higgs theory to be degenerated. The condition $V(\phi) = 0$
means that $|\phi| = a$, i.e., the set of vacuum values of the Higgs field
forms a sphere $S^2_{\rm vac}$ of radius $a$ in $d=3$ isotopic space. All the
points on this sphere are equivalent because there is a well defined $SU(2)$
gauge transformation which connects them.  

The solutions of the classical field equations map the vacuum manifold ${\cal
M} = S^2_{\rm vac}$ onto the boundary of 3-dimensional space, which is also a
sphere $S^2$.  These maps are charactered by a {\it winding number} $n = 0,\pm
1,\pm 2\dots$ which is the number of times $S^2_{\rm vac}$ is covered by a
single turn around the spatial boundary $S^2$.  The crucial point is that the
solutions having a finite energy on the spatial asymptotic could be separated
into different classes according to the behaviour of the field $\phi ^a$. The
trivial case is that the isotopic orientation of the fields do not depend
on the spatial coordinates and asymptotically the scalar field tends to the
limit
\begin{equation}                                 \label{triv}
\phi ^a = (0, 0, a)
\end{equation}
This situation corresponds to winding number $n=0$. 

We can also consider another type of solutions with the property that the
direction of isovector and isoscalar fields in isospace are functions of the
spatial coordinates. One could suppose that since the absolute minimum of the
energy corresponds to the trivial vacuum, such configurations would be
unstable.  However, the stability of them will be secured by the topology,
if we try to deform the fields continuously to the trivial vacuum
(\ref{triv}), then the energy functional would tend to infinity. In other
words, all the different topological sectors are separated by infinite
barriers.
  
To construct the solutions corresponding to the non-trivial minimum of the
energy functional (\ref{En}), we again consider the scalar field on the
spatial asymptotic $r \to \infty$ taking values on the vacuum manifold
$|\phi| = a$. However, we suppose that the isovector of the scalar field now
is directed in the isotopic space along the direction of the radius vector on
the spatial asymptotic\footnote{In the pioneering paper by Polyakov
\cite{Polyakov74} this solution was coined a `hedgehog'.}
\begin{equation}                                     \label{heudg}
\phi^a \stackrel{r\to \infty}{\longrightarrow} \frac{a r^a}{r}
\end{equation}
This asymptotic behaviour obviously mixes the spatial and isotopic indices and
defines a single mapping of the vacuum ${\cal M} $ onto the spatial
asymptotic, a single turn around the boundary $S^2$ leads to a single closed
path on the sphere $S^2_{\rm vac}$ and the winding number of such a mapping is
$n=1$.
 
As was mentioned by `t~Hooft \cite{Hooft74}, the configurations which are
characterised by different winding numbers cannot be continuously deformed
into each other. Indeed, the gauge transformation of
the form $U = e^{i ({\sigma}_k {\hat{\varphi}_k})\theta/2}$ rotates the
isovector to the third axis.  However, if we try to `comb the hedgehog', that
is, to rotate the scalar field everywhere in space to a given direction (so
called {\it unitary} or {\it Abelian gauge}), the singularity of the gauge
transformation on the south pole does not allow to do it globally. Thus, there
is no well-defined global gauge transformation which connects the
configurations (\ref{triv}) and (\ref{heudg}) and this singularity results in
the infinite barrier separating them.

\subsection{Definition of magnetic charge}

The condition of vanishing covariant derivative of the scalar field on the
spatial asymptotic (\ref{higgs-vacuum}) together with the choice of the
nontrivial hedgehog configuration implies that at $r \to \infty$
\begin{equation}
\partial_n\left(\frac{r^a}{r}\right) 
- e \varepsilon_{abc} A_{n}^b \frac{r^c}{r} = 0.
\end{equation}
The simple transformation
$$
\partial_n\left(\frac{r^a}{r}\right) 
= \frac{r^2 \delta_{an} - r_ar_n}{r^3} 
= \frac{1}{r}\left(\delta_{an}\delta_{ck} 
- \delta_{ak}\delta_{nc}\right) \frac{r_cr_k}{r^2} 
= -\varepsilon_{abc}\varepsilon_{bnk}\frac{r_cr_k}{r^3}
$$
then provides an asymptotic form of the gauge potential 
\begin{equation}                \label{A-as}
A_k^a (r) 
\stackrel{r\to \infty}{\longrightarrow} 
\frac{1}{e} \varepsilon_{ank}\frac{r_n}{r^2}
\end{equation}
This corresponds to the non-Abelian magnetic field
\begin{equation}
B_n^a  \stackrel{r\to \infty}{\longrightarrow} \frac{r_ar_n}{er^4}
\end{equation}
Therefore, the boundary conditions (\ref{heudg}), (\ref{A-as}) are compatible
with the existence of a long-range gauge field associated with an Abelian
subgroup which is unbroken in the vacuum. Since this field falls off like
$1/r^2$, which characterises the Coulomb-like field of a point charge, and
since the electric components of the field strength tensor (\ref{F}) vanish,
this configuration with a finite energy could be identified with a monopole.

To prove it, we first have to define the electromagnetic field strength
tensor. Recall that the unbroken electromagnetic subgroup $U(1)$ is associated
with rotations about the direction of the isovector $\phi$. Thus it would be
rather natural to introduce the electromagnetic potential as a projection of
the $SU(2)$ gauge potential $A_\mu^a$ onto that direction, see
eq.~(\ref{A-em}).  Furthermore, as was mentioned in the paper \cite{Corr76}, a
general solution of the equation $D_\mu\phi^a = 0$, for $\phi^a\phi^a=a^2$,
can be written as
\begin{equation}            \label{A-general}
A_\mu^a = \frac{1}{a^2e}\varepsilon_{abc}\phi^b\partial_{\mu}\phi^c 
+ \frac{1}{a} \phi^a \Lambda_\mu
\end{equation}
where $\Lambda_\mu$ is an arbitrary 4-vector. It can be identified with the
electromagnetic potential because eq.~(\ref{A-general}) yields for
$\phi^a\phi^a=a^2$:
$$
\frac{\phi^a}{a}A_\mu^a = \Lambda_\mu \equiv A^{\rm em}_\mu.
$$
Inserting eq.~(\ref{A-general}) into the definition of the field strength
tensor (\ref{F}) yields
\begin{equation}          \label{t-Hooft_F}
F_{\mu\nu}^a =  F_{\mu\nu}\frac{\phi^a}{a}, \qquad {\rm where} \qquad 
F_{\mu\nu} = \partial_\mu A_\nu - \partial_\nu A_\mu + \frac{1}{a^3e}
\varepsilon_{abc}\phi^a \partial_\mu\phi^b \partial_\nu\phi^c
\end{equation}

This gauge-invariant definition of the electromagnetic field strength tensor
$F_{\mu\nu}$ suggested by 't~Hooft \cite{Hooft74}, has a very deep meaning.
It is rather obvious that in the topologically trivial sector (\ref{triv}) the
last term in eq.~(\ref{t-Hooft_F}) vanishes and then we have
$$
F_{\mu\nu} =  \partial_\mu A_\nu - \partial_\nu A_\mu.
$$  
This is precisely the case of standard Maxwell electrodynamics.  Of course, in
this sector there is no place for a monopole because the Bianchi identities
are satisfied: $\partial^{\mu}{\widetilde F}_{\mu \nu} \equiv 0$.

However, for the configuration with non-trivial boundary conditions
(\ref{heudg}), (\ref{A-as}), also the Higgs field gives a non-vanishing
contribution to the electromagnetic field strength tensor
(\ref{t-Hooft_F}). Then, the second pair of Maxwell equations becomes modified:
\begin{equation}
\partial^{\mu}{\widetilde F}_{\mu \nu} = k_{\nu}
\end{equation}
Note, that if the electromagnetic potential $A_{\mu}^{\rm em}$ is regular, the
magnetic, or topological current $k_{\mu}$ is expressed via the scalar field
alone
\begin{equation}                     \label{top-current}   
k_\mu = \frac{1}{2}\varepsilon_{\mu\nu\rho\sigma}\partial^{\nu}F^{\rho\sigma}
=  \frac{1}{2a^3e}\varepsilon_{\mu\nu\rho\sigma}
\varepsilon_{abc}\partial^{\nu}\phi^a \partial^\rho\phi^b 
\partial^\sigma\phi^c.
\end{equation}
From the first glance this current is independent of any property of the
gauge field. It is conserved by its very definition:
\begin{equation} 
\partial_\mu k_\mu \equiv 0 \ ,
\end{equation}
unlike a Noether current which is conserved because of some symmetry.

Now we can justify the definition of magnetic charge
\cite{ArFreud}. According to eq.~(\ref{top-current})
\begin{eqnarray}                  \label{g-deff}
g &=& \int d^3x k_0 
= \frac{1}{2 e a^3}\int d^3x\,
\varepsilon_{abc}\varepsilon_{mnk}\,\partial_m\left(
\phi^a \partial_n\phi^b \partial_k\phi^c\right)\nonumber\\ 
&=& 
-\frac{1}{2 e a^3}\int d^2S_n\, \varepsilon_{abc}\varepsilon_{mnk}\,
\phi^a \partial_n\phi^b \partial_k\phi^c \ ,
\end{eqnarray}
where the last integral is taken over the surface of the sphere $S^2$ on the
spatial asymptotic. One can parameterise it by local coordinates
$\xi_\alpha,\, \alpha = 1,2$. Then we can write
\begin{equation}
\partial_n\phi^a = \frac{\partial\xi^\alpha}{\partial r^n} 
\frac{\partial \phi^a}{\partial \xi^\alpha}; \qquad 
d^2S_n = \frac{1}{2} \,
\varepsilon_{nmk}\frac{\partial r^m}{\partial \xi^\alpha} 
\frac{\partial r^k}{\partial \xi^\beta} \varepsilon_{\alpha\beta}d^2\xi.
\end{equation}
After some simple algebra we arrive at
\begin{equation}   \label{ch-monopole}
g = \frac{1}{2 e a^3}\int d^2 \xi\, 
\varepsilon_{\alpha\beta}\varepsilon_{abc}\,
\phi^a \partial_\alpha\phi^b \partial_\beta\phi^c
= \frac{1}{ e}\int d^2 \xi {\sqrt {\sf g}}  \ ,
\end{equation}
where ${\sf g} = \det(\partial_\alpha{\hat\phi}^a 
\partial_\beta {\hat \phi}^a)$ is the determinant of the metric tensor 
on the $S^2_{\rm vac}$ sphere in isospace. The magnetic charge is proportional
to an integer $n$ which mathematicians refer to as the Brouwer degree. The
geometrical interpretation of this integer is clear: it is the number of times
the isovector $\phi^a$ covers the sphere $S^2_{\rm vac}$ while $r^a$ covers
the sphere $S^2$ on the spatial asymptotic once. Thus \cite{ArFreud}:
\begin{equation}        \label{e=ng-non-abel}
g = \frac{4\pi n}{e}, \qquad n \in \mathbb{Z} \ ,
\end{equation}
where the factor $4 \pi$ is due to integration over the unit sphere.
This is the non-Abelian counterpart of the Dirac charge quantisation condition.

Another remark about the definition of the magnetic charge is that the Brouwer
degree and homotopic classification are equivalent to the 3d Poincar{\'e}--Hopf
index \cite{ArFreud}.  The latter is defined as a mapping of a sphere $S^2$
surrounding an isolated point $r_0$ where the scalar field vanishes, i.e.,
$\phi(r_0) = 0$, onto a sphere of unit radius $S^2_\phi$ (see e.g. \cite{Milnor}).
In other words, the
magnetic charge of an arbitrary field configuration can be defined as a sum of
the Poincar{\'e}--Hopf indices $i$ of non-degenerated zeros $r_0^{(k)}$ of the
Higgs field:
\begin{equation}
\label{3dindex}
g = \frac{4\pi}{e} \sum_k i ({r_0^{(k)}}) \ .
\end{equation}
Indeed, if we consider a scalar field which is constant everywhere in space
and satisfies the boundary condition (\ref{triv}), it has no zero at
all. Thus, the 3d Poincar{\'e}--Hopf index, alias magnetic charge, is equal to
zero. However, in the case of the hedgehog configuration which satisfies the
boundary condition (\ref{heudg}) 
\begin{equation}      
\phi^a = r^a h(r) \ ,
\end{equation}
where $h(r)$ is a smooth function having no zeros, there is a single zero at
the origin. Thus, $i(0)=1$ and that is a configuration of unit magnetic
charge.

This approach allows one to identify monopoles according to the positions of
zeros of the Higgs field. Such an identification is very useful from the point
of view of constructing multi-monopole solutions which we will consider in the
following section. But first we have to find a solution of the field
equations (\ref{Yang-Mills}), which would satisfy the boundary conditions
(\ref{heudg}) and (\ref{A-as}).

\subsection{'t~Hooft--Polyakov ansatz}

We showed that asymptotically, the monopole field configuration must satisfy
the conditions (\ref{heudg}) and (\ref{A-as}). Now, we try to define the
structure functions which form the radial shape of the monopole. As usual,
this problem can be simplified if we take into account the constraints
following from the symmetries of the configuration. Note, that we consider
static fields. That condition leaves only rotational $SO(3)$ symmetry from the
original Poincar\'e invariance of the Lagrangian (\ref{LagrGG}). Therefore,
the full invariance group of the system is $SO(3)\times SU(2)$, the product of
spatial and group rotations.  Moreover, the non-trivial asymptotic of the
Higgs field (\ref{heudg}) corresponds to the symmetry with respect to the
transformation from the diagonal $SO(3)$ subgroup which mix spatial and
group rotations. Thus, one can make the ansatz
\cite{Hooft74,Polyakov74}:
\begin{equation}                                 \label{Pola}
\phi ^a = \frac{r^a}{e r^2} H (aer); \quad A_n^a 
=  \varepsilon _{amn} \frac{r^m }{ er^2} [1 - K (aer)]; \quad A_0^a = 0 \ ,
\end{equation}
where $H (aer)$ and $K (aer)$ are functions of the dimensionless variable $\xi
= aer$. The explicit forms of these shape functions of the scalar and gauge
field can be found from the field equations. However, it would be much more
convenient to make use of the condition that the monopole solution corresponds
to a local minimum of the energy functional.  Substituting the ansatz
(\ref{Pola}) back into eq.~(\ref{En}) we have
\begin{eqnarray}                                           \label{mass}
E &=& \frac{4 \pi a }{e} \int \limits _0^{\infty }\frac{d \xi }{ \xi ^2} 
\biggl[
\xi ^2 \left( \frac{d K}{d \xi} \right) ^2 
+ \frac{1}{ 2}\left(\xi \frac{d H }{ d \xi} - H \right) ^2 \nonumber\\
& & 
\phantom{\frac{4 \pi a }{e} \int \limits _0^{\infty }\frac{d \xi }{ \xi ^2}}
+ \frac{1}{ 2} \left( K^2 - 1 \right)^2 + K^2 H^2 +
\frac{\lambda }{ 4e^2} \left(H ^2 - \xi ^2\right) ^2 \biggr] \ .
\nonumber
\end{eqnarray}
Variations of this functional with respect to the functions $H$ and $K$ yields 
\begin{equation}                                    \label{eqv}
\xi ^2 \frac{d^2 K}{ d \xi ^2} = K H^2 + K(K^2 -1); \qquad
\xi ^2 \frac{d^2 H}{ d \xi ^2} 
= 2K ^2 H + \frac{\lambda }{ e^2} H (H^2 -\xi ^2) \ .
\end{equation}
The functions
$K$ and $H$ must satisfy the following boundary conditions:
\begin{eqnarray}               \label{bound}
K(\xi) \rightarrow 1,\qquad H(\xi) \rightarrow 0 \quad {\rm for}\quad \xi 
\rightarrow 0 \ ;
\nonumber\\
K(\xi) \rightarrow 0,\qquad H(\xi) \rightarrow \xi \quad {\rm for}\quad \xi 
\rightarrow \infty \ ,
\end{eqnarray}
which correspond to the asymptotic (\ref{heudg}) and (\ref{A-as}). Indeed,
the substitution of the ansatz (\ref{Pola}) into the expressions for the
covariant derivative of the scalar field and the non-Abelian magnetic field
yields
\begin{eqnarray}              \label{D-cov}
D_n\phi^a &=& \frac{\delta_{an}}{er^2} KH + \frac{r^ar^n}{er^4}\left(
\xi\frac{d H}{d \xi}  - H - KH\right) 
\stackrel{r\to \infty}{\longrightarrow} 0;\nonumber\\
B_n^a &=& \frac{r_nr^a}{er^4} \left(1 - K^2 
+ \xi \frac{d K}{d \xi}\right)
- \frac{\delta_{an}}{er^2} \xi \frac{d K}{d \xi}
\stackrel{r\to \infty}{\longrightarrow} \frac{r_nr^a}{er^4} \ ..
\end{eqnarray}

Let us note that in the Higgs vacuum $D_n\phi^a =0$ and the electromagnetic
field strength is $F_{\mu\nu} = \phi^a F_{\mu\nu}^a/a$. Clearly, the magnetic
charge could be calculated as an integral over the surface of the sphere $S^2$
on spatial infinity (compare with eq. (\ref{g-deff})):
\begin{equation}        \label{g-integral}
g = \frac{1}{a} \int d^2S_n B_n 
= \frac{1}{a}\int  d^2S_n B_n^a \phi^a  
= \frac{1}{a}\int d^3x B_n^a D_n\phi^a  \ ,
\end{equation}
where we made use of the Bianchi identity for the tensor of non-Abelian
magnetic field $D_nB_n^a = 0$. Substituting the ansatz (\ref{Pola}) we obtain
\begin{eqnarray}                          \label{charge-g}
g &=& \frac{4\pi}{e} \int\limits_0^\infty \frac{d\xi}{\xi^2} \left\{
 (K^2-1)(H - \xi H^\prime) - 2\xi K^\prime KH \right\}\nonumber\\
&=& \frac{4 \pi}{e} \int\limits_0^\infty d\xi \frac{d}{d\xi}\left\{
\frac{1-K^2}{\xi}\right\} = \frac{4 \pi}{e} \ .
\end{eqnarray}   
Again we see that the boundary conditions (\ref{bound}) correspond to a
monopole of unit magnetic charge.

Numerical solutions of the system  (\ref{eqv}) were discussed in the papers  
\cite{BaisPrimak76,Kirkman81}. It turns out that the shape functions 
$H(\xi)$ and $K(\xi)$ approach rather fast to the asymptotic values. 
Thus, there is a Higgs vacuum outside of some region of the order of the
characteristic scale $R_c$, which is called the {\it core} of the
monopole. One could estimate this size by simple arguments
\cite{Preskill84}.  The total energy of the monopole configuration consists
of two components: the energy of the Abelian magnetic field outside the core
and the energy of the scalar field inside the core:
$$
E = E_{\rm mag} + E_{s} \sim 4\pi g^2 R_c^{-1} + 4\pi a^2 R_c \sim
\frac{4\pi}{e^2} \left(R_c^{-1} + m_v^2 R_c\right) \ .
$$
This sum is minimal if $R_c \sim m_v^{-1}$. In other words, inside the core at
distances shorter than the wavelength of the vector boson $m_v^{-1} \sim
(ae)^{-1}$, the original $SU(2)$ symmetry is restored. However, outside the
core this symmetry is spontaneously broken down to the Abelian electromagnetic
subgroup. In this sense there is no difference between the `t~Hooft--Polyakov
and the Dirac monopole outside the core \cite{GoddOlive}.

Unfortunately, the system of non-linear coupled differential equations
(\ref{eqv}) in general has no analytical solution. The only known exception is
the very special case $\lambda = 0$ \cite{Bog76,PS75,ColPras77}. This is the
so-called {\it Bogomol'nyi--Prasad--Sommerfield (BPS) limit} which deserves a special
consideration. 

\subsection{BPS monopole}
In the BPS limit of vanishing Higgs potential the 
scalar field also becomes massless and the energy 
of the static field configuration is taking the form
\begin{equation}
E  = \int\left\{ \frac{1}{4} \Tr\left(
\left(\varepsilon_{ijk} F_{ij}\pm D_k \Phi \right)^2 \right)
 \mp\frac{1}{2}\varepsilon_{ijk} \Tr\left( F_{ij} D_k \Phi \right)	
	\right\} d^3 r \ .
\label{E2}
\end{equation}
Thus, the absolute minimum of the energy corresponds to 
the static configurations which 
satisfy the first order Bogomol'nyi equations \cite{Bog76,PS75,ColPras77}:
\begin{equation}
\varepsilon_{ijk} F_{ij} = \pm  D_k \Phi 
\label{BPS}
\end{equation}
Substitution of the  `t~Hooft--Polyakov ansatz (\ref{Pola}) yields
the system of coupled differential equations of first order
\begin{equation}                      \label{BPS-str}
\xi \frac {dK}{d \xi} 
= - K H; \qquad \xi \frac {dH}{d \xi} 
= H + (1 - K^2) \ ,
\end{equation}
which have an analytical solution in terms of elementary functions:
\begin{equation}                      \label{BPS-solu}
K = \frac{\xi}{\sinh \xi }; \qquad H = \xi \coth \xi - 1 \ .
\end{equation}  
Definitely,  the solution to the first order BPS equation (\ref{BPS})
automatically satisfies the system of field equations of the second order,
(\ref{Yang-Mills}). 

As was mentioned in \cite{Ward81}, the BPS equation together with the Bianchi
identity means that $D_nD_n\phi^a = 0$, which precisely corresponds to the
field equation (\ref{Yang-Mills}). Therefore, the condition
$$
D_n\phi^a D_n\phi^a = (\partial_n\phi^a)(\partial_n\phi^a) + \phi^a 
(\partial_n\partial_n \phi^a) =
\frac{1}{2}\partial_n\partial_n(\phi^a\phi^a)
$$
holds.  The energy of the monopole configuration in the BPS limit is
independent from the properties of the gauge field and completely defined by
the Higgs field alone:
\cite{Ward81,Sutcliffe98}:
\begin{equation}
\begin{split}
E &= \frac{1}{2} \int d^3 x \partial_n \partial_n (\phi^a \phi^a) =
\frac{4\pi a}{e} \int \limits_0^\infty d\xi \frac{d}{d\xi}\left[ 
\xi H \frac{d}{d\xi} \left( \frac{H}{\xi} \right)\right]\nonumber\\
&= \frac{4\pi a}{e}\left( \coth \xi - \frac{1}{\xi}\right)
\left(1 - \frac{\xi^2}{\sinh^2 \xi}\right)
{\bigl| \bigr.}_{0}^\infty = \frac{4\pi a}{e} \ . 
\end{split}
\end{equation}
In comparison with the `t~Hooft--Polyakov solution, the behaviour of the Higgs
field of the monopole in the BPS limit has changed drastically. As we can see
from (\ref{BPS-solu}), alongside with the exponentially decaying component it
also obtains a long-distance Coulomb tail
\begin{equation}                  \label{as-higgs}
\phi^a \rightarrow a {\hat r}^a - \frac{r^a}{er^2}\quad {\rm for}\quad 
r \rightarrow \infty
\end{equation}
The reason for this is that in the limit $V(\phi)=0$ the scalar field becomes
massless.  Because an interaction, which is mediated by a massless scalar
field, always leads to attraction, the picture of the interaction between the
monopoles is very different in the BPS limit, as compared with the naive
picture based on pure electromagnetic interaction. Manton showed that  
monopole--monopole interaction is composed of two 
parts originated from the long-range scalar force and the standard
electromagnetic interaction, which could be either attractive or repulsive
\cite{Manton77}. Mutual compensation of both contributions leaves the pair of
BPS monopoles static but the monopole and anti-monopole interact with
double strength.

Many of the remarkable properties of the BPS equation (\ref{BPS}) are
connected with its property of integrability. As was pointed out by Manton
\cite{Manton78}, integrability of the BPS system is connected with a one-to-one
correspondence between the system of BPS equations and the reduced equations
of self-duality of the pure Euclidean Yang--Mills theory.  Indeed, the
Julia--Zee correspondence means that
\begin{eqnarray}                                                   \label{su2}
D_n\phi^a &\rightleftharpoons& D_nA_0^a \equiv F_{0n}^a, \nonumber\\
B_n^a = D_n\phi^a &\rightleftharpoons& {\widetilde F}_{0 n} = F_{0n}^a \ .
\end{eqnarray}
Therefore, if we suppose that all the fields are static, the Euclidean
equations of self-duality $F_{\mu\nu}^a = {\widetilde F}_{\mu \nu}^a$ reduces
to the equations (\ref{BPS}) and the monopole solutions in the Bogomol'nyi limit
could be considered as a special class of self-dual fields.

Of course, it would not be quite correct to make a direct identification
between these fields and instantons, because the instanton configuration could
be independent from Euclidean time only in the limit of infinite
action. Nevertheless, this analogy opens a way to apply in the $d=3+1$
monopole theory the same very powerful methods of algebraic and differential
geometry which were used to construct multi-instanton solutions of the
self-duality equations in $d=4$ \cite{AtHit}. In particular, in the case of
the BPS monopole, the solution of the self-duality equations could be
constructed on the ansatz of Corrigan and Fairlie \cite{Corrigan77}.

The analogy between the Euclidean Yang--Mills theory and the BPS equations
could be traced up to the solutions. It was shown \cite{Rossi79,Rossi82}, that
the solutions of these equations exactly equal to an infinite chain of
instantons directed along the Euclidean time axis $t$ in $d=4$.

 It has also
been shown by Manton \cite{Manton78}, that such a multi-instanton
configuration can be written on the `t~Hooft ansatz with the help of a
superpotential $\rho(r,t)$ as:
\begin{equation}                              \label{multi-inst}
A_n^a = \varepsilon_{anm}\partial_m \ln \rho 
+ \delta_{an} \partial_0 \ln \rho; \qquad 
A_0^a = -\partial_a \ln \rho \ ,
\end{equation}
where the sum over the infinite number of instantons is performed in the
superpotential:
$$
\rho = \sum\limits_{n=-\infty}^{n=\infty} \frac{1}{\xi^2 + (\tau -2\pi n)^2},
\quad {\rm where}\quad \xi = aer, \quad \tau = ae t \ .
$$ 
Here, the distance between the neighbouring instantons is equal to $2\pi$ in
units of $\tau$ and the size of the instanton is equal to one in units of
$\xi$.

Rossi pointed out \cite{Rossi79} that this sum over instantons could be
calculated analytically. Indeed, the superpotential can be decomposed into two
sums over Matsubara frequencies $\omega_n = 2\pi n$ which are well known from
statistical physics:
\begin{equation}
\rho =\frac{1}{2\xi}\left\{ \sum\limits_{n=-\infty}^{n=\infty}
\frac{1}{\xi + i\tau - 2i\pi n}  
+ \sum\limits_{n=-\infty}^{n=\infty}
\frac{1}{\xi - i\tau + 2i\pi n}\right\}
\end{equation}
Introducing the complex variable $z = \xi + i\tau$, we can write
\begin{eqnarray}
\rho &=&\frac{1}{2\xi}\left\{ \sum\limits_{n=-\infty}^{n=\infty}
\frac{1}{z  - i\omega_n}  
+ \sum\limits_{n=-\infty}^{n=\infty}
\frac{1}{z^* + i\omega_n}\right\} 
= \frac{1}{2\xi}\left\{ \coth \frac{z}{2} + \coth \frac{z^*}{2} \right\}
\nonumber\\
&=& \frac{1}{2\xi} \frac{\sinh \xi}{\cosh \xi - \cos \tau} \ .
\end{eqnarray}

Substitution of this result into the potential (\ref{multi-inst}) corresponds
to the ``dyon in the `t Hooft gauge''.  This solution is periodic in
time. However, the time-dependent periodic gauge transformation of the form
\cite{Manton78}
\begin{equation}
U = \exp\left\{\frac{i}{2a}{\hat r}^a \sigma^a \omega\right\},
\qquad {\rm where}\quad 
\tan \omega = \frac{\sin \tau \sinh\xi}{\cosh\xi \cos \tau -1}
\end{equation}
transforms the infinite chain of instantons (\ref{multi-inst}) into the form:
\begin{equation}     \label{BPS-sol}
A_n^a = \varepsilon_{anm} 
\frac{r^m}{er^2}\left(1 - \frac{\xi}{\sinh \xi}\right);
\qquad
A_0^a = a {\hat r}^a\left(\coth \xi - \frac{1}{\xi}\right) \ .
\end{equation}
This is exactly the monopole solution of the BPS equation (\ref{BPS-solu}) but
with the time component of the gauge potential replacing the scalar
field. This is the so called ``dyon in the Rossi gauge''. Thus, the Julia--Zee
correspondence establishes an exact relation between a single BPS monopole and
an infinite instanton chain.

As mentioned above, the action of the infinite number of instantons is
divergent:
\begin{equation}
S = \sum_n S_1 = \sum_n\frac{8\pi^2 n}{e^2} \to \infty \ .
\end{equation}
However, the mass of the monopole being defined as an action per unit of
Euclidean time is, of course, finite \cite{Rossi79}:
\begin{equation}
\frac{dS}{dt} = \frac{8\pi^2}{e^2}\frac{ae}{2\pi} = \frac{4\pi a}{e} \equiv M \ .
\end{equation}   
To sum up, the BPS monopole is equivalent to an infinite chain of instantons
having identical orientation in isospace and separated by an interval
$\tau_0 = 2\pi$. An alternative configuration is a chain of correlated
instanton--anti-instanton pairs, which corresponds to an infinite monopole
loop.

\section{Multimonopoles}

So far, we have considered a single static monopole which has topological
charge $n=1$.  An obvious generalisation would be a solution of the field
equation with an arbitrary integer topological charge $n$. At this stage we
have to consider two possibilities: a single `fat' (possibly unstable!)
monopole, having a charge $n>1$, or a system of several monopoles with a total
charge equal to $n$. Obviously, the second situation could be much more
interesting, because in this case one would encounter the
effects of interaction between the monopoles, their scattering and decay.

Over the last 20 years, the investigation of the exact multi-monopole
configurations was definitely at the crossing of the most fascinating
directions of modern field theory and differential geometry.  This kind of
research may have caused more enthusiasm from the side of the mathematicians,
rather than the physicists. The point is that the Bogomol'nyi equation is 
3-dimensional reduction of the integrable self-duality equations. Thus, its
solution is simpler than the investigation of the multi-instanton
configurations which arise in Yang--Mills theory in $d=4$. On the other hand, 
there is a significant interest to the BPS configurations in supersymmetric 
theories and relation of these objects to electro-magnetic duality.

The property of integrability makes it possible to find the complete set of
multi-monopole solutions. This, however, requires sophisticated mathematical
techniques, for example the Atiyah--Drinfeld--Hitchin--Manin (ADHM)
construction and other methods
developed over the last years.  Of course, any attempt to give a detailed
description of this fast developing and very intriguing subject is outside the
scope of this brief review.  We consider here only some elementary
aspects of the multi-monopoles. 

\subsection{BSP multimonopoles: bird's eye view}
\label{secBPS}
The exact cancellation of the electromagnetic attraction and the dilaton
repulsion in the two-monopole system suggests to conjecture that there are
multi-monopole static solutions of the Bogomol'nyi equations. Since Bogomol'nyi
has found that the explicit spherically symmetric solution with $n = 1$ is
unique \cite{Bog76}, any possible multi-monopole configuration with $n > 1$
cannot have such a symmetry. Thus, the structure of the 
configuration we have to deal with, is rather complicated and it is difficult to
obtain these solutions. 
 
Naively, one can visualise the following geometric transformation which could
help to construct a multi-monopole configuration starting from a given single
spherically symmetric BPS monopole. Recall that magnetic charge is associated
with the asymptotic behaviour of the scalar field which form a sphere $S^2$ for
a monopole of charge $n=1$. To construct a 2-monopole configuration we shall
remove from this sphere the equatorial circle $S^1$ as shown in
figure~\ref{f-002}, and then identify all the points on the
equators of the two hemispheres with the north and south poles of
two new spheres respectively. Construction of an $n$-monopole
configuration requires a simple iteration of this procedure\footnote{With
some imagination one can compare this picture with the well known process of
biological cell division...}.
\begin{figure}[thb]
\begin{center}
\setlength{\unitlength}{1cm}
\lbfig{f-002}
\begin{picture}(13,2.7)
\put(-0.3,-0.2)
{\mbox{
\psfig{figure=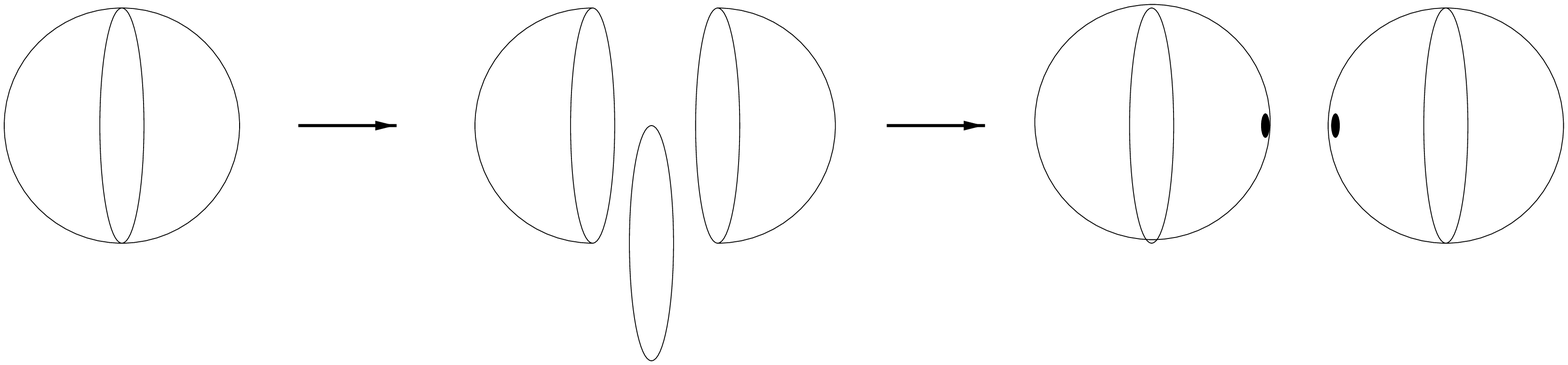,height=3.0cm}}} 
\end{picture} 
\caption{
Geometry of the transformation which allows to construct 
two separated monopoles starting from spherically symmetric one-monopole
configuration.} 
\end{center} 
\end{figure} 
However, an adequate mathematical description of the geometrical
transformation which could solve the multi-monopole problem is not so trivial
and emerged only as result of heavy work of many mathematicians lasted
over a decade. The problem is not to construct an explicit solution for an
arbitrary $n$-monopole configuration but to prove the completeness of the
solution, i.e., to prove that all possible solutions are generated by this
procedure.
 
To give some clue which methods we have to use, let us note that the naive
picture above is closely connected with the mathematical apparatus of
projective geometry.  Indeed, making a standard stereographic projection of
the sphere $S^2$ onto a plane, we see that removing the equator of the sphere
corresponds to a cut on the projective plane which then becomes isomorphic to
a doubly covered Riemann surface.  Recall also that identification of the
antipodal points of a sphere $S^n$ transform it into a real projective space
$\mathbb{R} P^n$. Thus, it is no surprise that the powerful methods of twistor geometry 
were fully exploited to construct the multi-monopole configurations.

Generally speaking, to construct a general explicit $n$-monopole solution, one
has to make use of the integrability of the Bogomol'nyi equation. There are
three different approaches to this problem which use:
\begin{enumerate}
\item
twistor technique (the so called Atiyah--Ward ansatz
\cite{AtWard77,Corr78,Ward81,Corr81});
\item 
Atiyah--Drinfeld--Hitchin--Manin (ADHM) construction \cite{ADHM78}, which
was modified by Nahm \cite{Nahm82};
\item 
inverse scattering method (Riemann-Hilbert problem), which was applied to
the linearised Bogomol'nyi equation \cite{Forcas80,Forcas81,Forcas82}.
\end{enumerate}

The reader wishing to know more about first and second directions 
should consult the classic book by Atiyah and Hitchin \cite{AtHit} and
the original papers \cite{AtHit85,At87,Ward81}. A comprehensive review of the
mathematical aspects of this problem can be found in \cite{Taubes}.  The very
detailed review by Nahm in the collection \cite{Craigie86} is essential
reading.  Recent developments are discussed in a very good review by Sutcliffe
\cite{Sutcliffe98}. For a detailed description of the last approach we refer 
the reader to the comprehensive review \cite{Forcas82}. Here, we briefly outline 
another development, related with recently discovered {\it essentially non-BPS} 
multimonopoles \cite{Rueber85,KKT,Sutcliffe99,KKS1,KKS2}.

\subsection{Singular $\bf SU(2)$ monopole with charge $\bf g = ng_0$} 

As we attempt to construct in a non-Abelian theory a system of several
(anti)monopoles, we run into a problem connected with the self-coupling of the
gauge fields and non-linearity of the field equations. A simple superposition
of the fields of two or more monopoles is no longer a stationary point of the
action. However, as we saw in the previous section, in the Yang-Mills-Higgs
non-Abelian model the gauge invariant electromagnetic field strength tensor
(\ref{t-Hooft_F}) is defined as
\begin{equation}                 \label{em-F}
F_{\mu\nu} = \partial_\mu A_{\nu} - \partial_{\nu}A_{\mu}
- \frac{1}{e}\varepsilon_{abc}{\hat \phi}^a \partial_{\mu} {\hat \phi}^b
\partial_\nu {\hat \phi}^c,
\end{equation}
where the abelian gauge potential is projected out as 
$A_\mu = A_\mu^a{\hat \phi}^a$. Thus, we can perform a singular gauge
transformations which rotates this configuration to an {\it Abelian gauge}
where the scalar field is constant: ${\tilde \phi}^a = a \delta_{a3}$ and the
gauge field has only one isotopic component, ${\tilde A}_k^3$. In such a gauge
the second term in the definition of the electromagnetic field strength tensor
(\ref{em-F}) transforms into the singular field of the Dirac string and the
magnetic charge is entirely associated with the topology of the gauge field 
\cite{ArFreud}.

Recall that gauge transformations, which connect the `hedgehog' and Abelian
gauges, are singular and we have to be very careful dealing with
them. However, there is an obvious advantage in working in the Abelian gauge:
here the gauge potentials are additive and the field equations are
linear. That is why the authors of the paper \cite{ArFreud} suggested to
implement the following program of construction of a multi-monopole
configuration: (i) start with an Abelian gauge, (ii) suppose that the gauge
potential ${\tilde A}_k^3$ is a simple sum of a few singular Dirac monopoles
embedded into the $SU(2)$ gauge group and then try to define a gauge
transformation, which
\begin{enumerate}
\item
Removes the singularity of the potential ${\tilde A}_k^3$ in the string gauge;
\item
Provides proper asymptotic behaviour of the fields in the Higgs vacuum: the
scalar field must smoothly tend to the vacuum value $|\phi| = a$, while the
gauge potential must vanish as $1/r$.
\end{enumerate}
  
Let us try to implement this program to construct a possible generalisation of
the `t~Hooft--Polyakov solution (\ref{Pola}) to the case of non-minimal
magnetic charge $g = ng_0 = 4\pi n/e$ \cite{Bais76,Fujii89,Frampton76}. 
We consider a `fat' Dirac potential  embedded 
into the $SU(2)$ model. Thus, we start with
the Abelian gauge where the electromagnetic subgroup corresponds to rotation
about the third axis of isospace:
\begin{equation}                  \label{nA}
{\tilde A}_k ({\bf r}) 
= \frac{n}{er}\frac{1 - \cos \theta}{\sin\theta}\, T_3 {\hat \varphi}_k \ .
\end{equation}
In the following discussion we will consider the fundamental representation
of the group, $T_a=\frac{\sigma_a}{2}$.

Now we can make use of an analogy with the gauge transformation 
which rotates the Dirac potential to the non-Abelian Wu--Yang singular
potential:
\begin{equation}
A_k = \frac{1}{2}{A^a_k}\sigma^a 
= {U}^{-1}{\tilde A}_k{U} -\frac{i}{e}{U}^{-1}\partial_k U, \qquad 
\phi^a = U{\tilde \phi}^a U^{-1} \ ,
\end{equation}
where
\begin{equation} \label{n-gauge}
U(\theta,\varphi) 
= e^{-i ({\bf \sigma}_k {\hat{\bf \varphi}_k}^{(n)})\theta/2} 
= \left( \begin{array}{cc} 
\cos \frac{\theta}{2} & -\sin \frac{\theta}{2} e^{-in\varphi} \\[3pt]
\sin \frac{\theta}{2} e^{in\varphi} & \cos \frac{\theta}{2}  
\end{array}\right) \ .
\end{equation}
Here the $n$-fold rotation in azimuthal angle $\varphi$ 
is needed to balance 
the singular part of the Abelian potential (\ref{nA}), 
$ \hat{\vecphi}^{(n)} = 
- {\hat {\bf e}}_1 \sin n\varphi + {\hat {\bf e}}_2 
\cos n\varphi$. 

To write the rotated potential in a compact form  
we define the $su(2)$ matrices 
$\tau_r^{(n)} = ({\hat{\bf r}}^{(n)} \cdot \vecsigma)$, 
$\tau_\theta^{(n)}= ({\hat \vectheta}^{(n)} \cdot \vecsigma) $, 
and $\tau_\vphi^{(n)} = ({\hat \vecphi}^{(n)} \cdot \vecsigma) $.
  In this notation we obtain 
\begin{equation}                  \label{n-singular}
A_k = \frac{1}{2er}\left(\tau_\varphi^{(n)} 
{\hat \theta}_k - n \tau_\theta^{(n)} {\hat \varphi}_k\right);
\qquad \phi^a = a {\hat r}_a^{(n)} \ .
\end{equation} 
As one could expect, when rotated into the `hedgehog' gauge, this
configuration is spherically symmetric and the corresponding magnetic field
\begin{equation} 
B_k = n \frac{r_k}{er^3}
\end{equation}
is exactly the field of a static magnetic monopole with charge $g=4\pi n/e$ at
the origin.  One can prove that for the configuration (\ref{n-singular}) the
condition $D_n\phi^a = 0$ holds. 

Note that the potential (\ref{n-singular}) is not a naive generalisation of 
the Wu-Yang potential  
\begin{equation}             \label{n-Wu-Yang}
A_n = n\, \varepsilon_{amn}\frac{r_m}{r^2} \ ,
\frac{\sigma^a}{2}
\end{equation}
which was considered in \cite{Zakharov01}. The configuration 
(\ref{n-Wu-Yang}) with $n=2$ has some amusing properties: The corresponding 
field strength tensor vanishes identically since the commutator term precisely 
cancel the derivative terms. Indeed, such a  potential is a pure gauge:
$$
A_n = i U^{-1} \partial_n U, \quad {\rm where} \quad U = i\sigma^a {\hat r}^a 
$$
unlike the potential  (\ref{n-singular}). Thus, this configuration is an unstable 
deformation of the topologically trival sector. We shall consider such deformations 
below. 

Recall that the Wu--Yang configuration  
which was constructed via the gauge rotation of the embedded Dirac potential,  
is only the asymptotic limit of the `t~Hooft--Polyakov solution 
(\ref{Pola}) at $r \to \infty$. Unlike the former, the latter corresponds to
finite energy of the configuration.  It is easy to see that the configuration
(\ref{n-singular}) is singular at the origin as well. 

One can try to exploit an analogy with the `t~Hooft--Polyakov ansatz, i.e.,
modify the asymptotic form of the fields (\ref{n-singular}) by including shape
functions $H(r)$ and $K(r)$, respectively \cite{Bais76,Frampton76}:
\begin{equation}                    \label{n-reg}
A_k = \frac{K(r)}{2er}
\left(\tau_\varphi^{(n)} 
{\hat \theta}_k - n \tau_\theta^{(n)} {\hat \varphi}_k\right); \qquad 
\phi^a = a H(r) {\hat r}_a^{(n)} \ .
\end{equation}
However, substitution of this ansatz into the field equations of the
Yang--Mills--Higgs system (\ref{Yang-Mills}) for $|n| \ge 2$ leads to a
contradiction with the assumption that a regular solution of the form
(\ref{n-reg}) could exist.  Thus, we need to introduce more profile functions 
to obtain a smooth non-spherically symmetric solution of the field equations. 
However even that configuration will be unstable.  
Here, we can see a manifestation of the very
general Lubkin theorem \cite{Lubkin63} (see, for example, the discussion in
the Coleman lectures \cite{Coleman81} and Nahm review in \cite{Craigie86}).
According to this theorem, there is a unique spherically symmetric monopole in
the $SU(2)$ model (\ref{LagrGG}) with minimal magnetic charge.  Both
analytical and numerical calculations have proved this conclusion
\cite{Brand-Neri79,Baacke92}. Therefore, the configuration, which has the 
asymptotic form (\ref{n-reg}), is a saddle point of the energy functional 
and it decays into a system of a few separated 
single monopoles and antimonopoles with total charge $n = n_+ - n_-$.  

Indeed, there are axially symmetric saddle point solutions of the model (\ref{LagrGG}) which 
approach the form (\ref{n-singular}) 
on the spacial asymptotic \cite{Rossi77}. We review these monopoles below, 
in section \ref{multisol}. 

\subsection{Magnetic dipole}
\label{dipole}

Let us continue our attempts to construct
multi-monopole configurations. We consider now 
a magnetic dipole: a monopole--anti-monopole
pair located on the $z$ axis at the points $(0,0,\pm L)$ with both strings
directed along the positive $z$ axis as in figure~\ref{f-01}.  A
simple addition of the corresponding two singular Dirac potentials yields
\begin{equation}                  \label{g-g}
{A}_k ({\bf r}) 
= \frac{1}{2e}
\left(\frac{1 - \cos \theta_1}{r_1 \sin\theta_1}
- \frac{1 - \cos \theta_2}{r_2 \sin\theta_2} \right) 
\sigma_3\, {\hat \varphi}_k \ .
\end{equation}

\begin{figure}[thb]
\begin{center}
\setlength{\unitlength}{1cm}
\lbfig{f-01}
\begin{picture}(13,6.2)
\put(1.8,-0.3)
{\mbox{
\psfig{figure=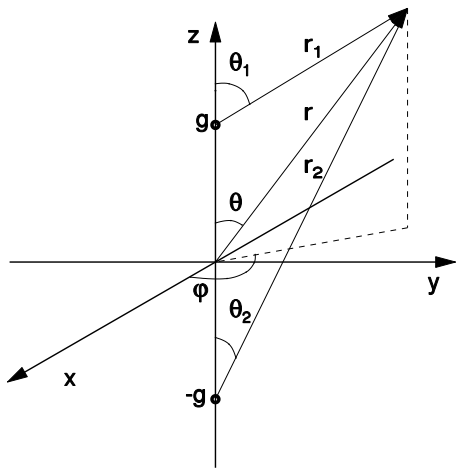,height=7.3cm}}} 
\end{picture} 
\caption{
Magnetic dipole configuration from afar.}
\end{center} 
\end{figure} 

It was noted by Coleman that a crucial feature of this construction is that
the Dirac strings of both monopoles lie along the same axis
\cite{Coleman81}.  That allows to define a gauge transformation which removes
the singularity of the potential (\ref{g-g}) embedded into the
$SU(2)$ group \cite{ArFreud,Bais76}.  Let us exploit the analogy with the
transition of a monopole embedded into $SU(2)$, from the singular Dirac
monopole in the Abelian gauge to the Wu--Yang non-Abelian monopole in the
`hedgehog' gauge. We see that the gauge transformation which would remove the
string singularity must rotate a unit isovector ${\hat {\bf n}}$ associated
with the direction of the string, about the third isospace axis by an angle
$4\pi$. However, this vector now originates from the points $| z |
=L$. Therefore, the gauge transformation we are looking for, must become an
element of unity for $| z | > L$. A proper choice is \cite{ArFreud,Bais76}
\begin{equation}                         \label{2-gauge} 
U(\theta_1 - \theta_2, \varphi)  
= e^{-\frac{i}{2} \varphi \sigma_3}  
e^{\frac{i}{2}(\theta_1 - \theta_2) \sigma_2}
e^{\frac{i}{2} \varphi \sigma_3}
= \left(\begin{array}{cc} 
\cos \frac{\theta_1 - \theta_2}{2} 
& -\sin \frac{\theta_1 - \theta_2}{2} e^{-i\phi} \\[3pt]
\sin \frac{\theta_1 - \theta_2}{2} e^{i\phi} 
& \cos \frac{\theta_1 - \theta_2}{2}  
\end{array}\right) \ .
\end{equation}
A simple calculation shows that the Higgs field after rotation into the
`hedgehog' gauge is
\begin{equation}                \label{multi-higgs}
{\tilde {\phi}^a} = U \phi^a U^{-1} 
= a (\sin (\theta_1-\theta_2) \cos \varphi,
\sin (\theta_1-\theta_2) \sin \varphi,
\cos (\theta_1-\theta_2))
\end{equation}
One can prove that the gauge transformation (\ref{2-gauge}) removes
the singularity of the embedded potential (\ref{g-g}).

A generalization of this procedure allows also to rotate into a non-singular
gauge some other configurations: a monopole--anti-monopole pair connected with
a Dirac string, a monopole--monopole pair, or a system of a few monopoles
lying along a line \cite{Bais76}.  Moreover, it is possible to generalize this
procedure to the case of an arbitrary gauge group, for example, an $SU(3)$
magnetic dipole was considered in \cite{Olesz87}. Recently we make use of the
Abelian gauge to analyse interaction between two different well separated non-BPS $SU(3)$ 
fundamental monopoles; this is an amusing situation of attraction between two 
positive or negative charges \cite{S03}.   
The only restriction is
that all the strings must be directed along the same line;
otherwise it is impossible to remove all the singularities of the
multi-monopole potential by making use of a singular gauge transformation
\cite{Coleman81}. In other words, such a multi-monopole configuration after
rotation to a `hedgehog' gauge is described as a system of a few monopoles
having identical orientation in isotopic space. Obviously, that is not the
case for an arbitrary multi-monopole system we are looking for.

Another inconsistency of the description above is that these expressions have
a restricted domain of applicability. 

Indeed, there was a hidden contradiction in our discussion
above. Actually, so far we are dealing 
with point-like monopoles because our configuration is just a generalization
of the non-Abelian Wu--Yang potential.  The regular `t~Hooft--Polyakov
solution coincides with it only asymptotically. Thus, there is still a
question of the inner structure of the monopoles, or in other words, the
problem of finding a solution which would make the Higgs field vanish at some
points, which are associated with the positions of the monopoles. 

The
contradiction is that, on one hand we suppose that each monopole is
characterized by a topological charge connected with the spatial asymptotic of
the scalar field. On the other hand, the definition of a magnetic dipole 
implies that the monopoles are separated by a finite
distance $2L$, and moreover, $L \ll r$. A proper approximation would therefore
not be a magnetic dipole, but rather a monopole--anti-monopole pair separated
by a distance which is very large compared to size of the core. 
Indeed, there is a smooth, finite
energy magnetic dipole solution to the model (\ref{LagrGG})
where two zeros of the Higgs field are relatively close to each other \cite{Rueber85,mapKK}.

\subsection{Rebbi-Rossi multimonopoles, monopole-antimonopole chains and 
closed vortices}
\label{multisol}

Discussion of the magnetic dipole `from afar', yields some clue to the 
structure of the solution we sought for. 
The configuration space of YMH theory consists of sectors characterized by the topological
charge of the Higgs field. 
While the unit charge  
`t Hooft - Polyakov hedgehog solution  (\ref{Pola}) corresponds 
to a single covering of the vacuum manifols $S^2_{\rm vac}$ is by a
single turn around the spatial boundary $S^2$, multimonopole configurations 
have to be characterised by $n$-fold covering of the vacuum manifold. 
This configuration is spherically symmetric.
It was shown that $SU(2)$ monopoles with higher topological 
charge cannot be spherically 
symmetric \cite{WeinbergGuth} and posess at most axial symmetry 
\cite{Rossi77,Ward81,Forgacs,Prasad} or no rotational symmetry at all \cite{Sutcliffe98,Sutcliffe99-1}. 

In the BPS limit of vanishing Higgs potential
spherically symmetric monopole and axially symmetric multimonopole solutions, 
which satisfy the first order Bogomol'nyi equations (\ref{BPS})
are known analytically  \cite{Ward81,Forgacs,Prasad}. These 
solutions were constructed on the way which 
we outlined in section \ref{secBPS}. 
For these solutions all zeros of the Higgs field are 
superimposed at a single point.  
Multimonopole solutions of the Bogomol'nyi equations
which do not possess any rotational symmetry \cite{CorGod},
have recently been constructed numerically by making use of the 
rational map ansatz \cite{Sutcliffe98}. In these solutions
the zeros of the Higgs field are no longer all
superimposed at a single point but are located at
several isolated points \footnote{Static spherically symmetric 
$SU(N)$ non-Bogomol'nyi BPS monopoles were recently constructed for 
$N>2$ \cite{Sutcliffe99}.}.

The energy of the BPS solutions satisfies exactly the lower energy 
bound given by the topological charge. As shown by Taubes using infinite dimensional 
Morse theory, in each topological sector 
there exist in addition smooth, finite energy solutions of the second order
field equations, which do not satisfy the Bogomol'nyi equations but only the second 
order field equations  (\ref{Yang-Mills}) \cite{Taubes82}. 
Consequently, the energy of these solutions exceeds
the Bogomol'nyi bound.

The simplest such solution resides in the  
sector with zero topological charge 
and corresponds to a saddlepoint of the energy functional
\cite{Taubes82}.   
It possesses axial symmetry, and the two zeros of its
Higgs field are located symmetrically on the positive and negative 
$z$-axis. This solution corresponds to a monopole and 
antimonopole in static equilibrium \cite{Rueber85,mapKK}. 
Such a configuration is a deformations of the topologically 
trivial sector. 

Spherically symmetric non-Bogomol'nyi BPS $SU(N)$ monopoles were 
considered by Ioannidou and Sutcliffe \cite{Sutcliffe99}. 
Very recently we discovered  new classical solutions, which are associated with monopole-antimonopole
systems in $SU(2)$ YMH theory with the scalar field in the adjoint 
representation \cite{KKS1,KKS2,KKS3}. In these solutions, which generalizes both Rebbi-Rossi 
multimonopoles \cite{Rossi77} and monopole-antimonopole pair solution \cite{Rueber85,mapKK},  
the Higgs field vanishes 
either at some set of discrete isolated points or at rings. The latter configurations corresponds 
to the closed vortices while the former are (multi)-monopole-antimonopole bound systems. There 
are also a third class of solutions which 
corresponds  to a single (multi) monopole bounded with a system of vortex rings centered around the 
symmetry axis. We review these configurations below.

\subsubsection{Static axially symmetric ansatz}
Since the Higgs field is taking values in $su(2)$ Lie algebra, we may 
consider a triplet of unit vectors 
\begin{equation}
\begin{split}
{\hat e}_r^{(n,m)} &= [\sin(m\theta) \cos(n\vphi), \sin(m\theta)\sin(n\vphi), \cos(m\theta)];\\
{\hat e}_\theta^{(n,m)} &= [\cos(m\theta) \cos(n\vphi), \cos(m\theta)\sin(n\vphi), -\sin(m\theta)];\\
{\hat e}_\vphi^{(n)} &= [-\sin(n\vphi), \cos(n\vphi), 0]
\label{unit_e}
\end{split}
\end{equation}
which describe both rotations in azimuthal angle and in polar angle as well. 

Now we define the $su(2)$ matrices
$\tau_r^{(n,m)}$, $\tau_\theta^{(n,m)}$, and $\tau_\vphi^{(n)}$ as a product 
of these vectors with 
the usual Pauli matrices $\tau^a = (\tau_x, \tau_y, \tau_z)$: 
\begin{eqnarray}
\tau_r^{(n,m)}  & = &
\sin(m\theta) \tau_\rho^{(n)} + \cos(m\theta) \tau_z \ ,
\nonumber\\
\tau_\theta^{(n,m)} & = &
\cos(m\theta) \tau_\rho^{(n)} - \sin(m\theta) \tau_z \ ,
\nonumber\\
\tau_\vphi^{(n)} & = &
 -\sin(n\vphi) \tau_x + \cos(n\vphi)\tau_y \ ,
\nonumber
\end{eqnarray}
where $\tau_\rho^{(n)} =\cos(n\vphi) \tau_x + \sin(n\vphi)\tau_y $ and $\rho = \sqrt{x^2 + y^2} = 
r\sin\theta$. 

We  parametrize the gauge potential and the Higgs field by the static, purely magnetic Ansatz
\begin{eqnarray}
A_\mu dx^\mu
& = &
\left( \frac{K_1}{r} dr + (1-K_2)d\theta\right)\frac{\tau_\vphi^{(n)}}{2e}
-n \sin\theta \left( K_3\frac{\tau_r^{(n,m)}}{2e}
                     +(1-K_4)\frac{\tau_\theta^{(n,m)}}{2e}\right) d\vphi
\label{ansatzA} \\
\Phi
& = &
\Phi_1\tau_r^{(n,m)}+ \Phi_2\tau_\theta^{(n,m)} \  .
\label{ansatzPhi}
\end{eqnarray}
which generalises the spherically symmetric `t Hooft-Polyakov ansatz 
(\ref{Pola}). The latter can be recovered if we impose the constrains 
$K_1=K_3=\Phi_2=0, K_2=K_4=K(\xi), \Phi_1 = H(\xi)$.

We refer to the integers $m$ and $n$ in (\ref{unit_e}),(\ref{ansatzA}),(\ref{ansatzPhi}) as 
$\theta$ winding number and $\vphi$ winding number, respectively. Indeed, 
as the unit vector (\ref{unit_e}) parametrized by the polar angle 
$\theta$ and asimuthal angle $\vphi$ covers the sphere $S_2$ once, the fields defined by the Ansatz 
(\ref{ansatzA}),(\ref{ansatzPhi}) wind $n$ and $m$ times around $z$-axis and $\rho$-axis respectively. 

There are some useful relations between the matrices 
$\tau_r^{(n,m)}$, $\tau_\theta^{(n,m)}$, and $\tau_\vphi^{(n)}$:
\begin{equation}
\begin{split}
\partial_\theta \tau_r^{(n,m)} = m \tau_\theta^{(n,m)};&\qquad 
\partial_\theta \tau_\theta^{(n,m)} = -m \tau_r^{(n,m)};\qquad 
\partial_\vphi \tau_r^{(n,m)} = n \sin (m\theta) \tau_\vphi^{(n)}\\
\partial_\vphi \tau_\theta^{(n,m)} = n \cos (m\theta) \tau_\vphi^{(n)};&\qquad 
\partial_\vphi \tau_\vphi^{(n)} = -n[\sin (m\theta) \tau_r^{(n,m)} + \cos (m\theta)\tau_\theta^{(n,m)}];\\
&\tau_z = \cos (m\theta) \tau_r^{(n,m)} - \sin (m\theta)\tau_\theta^{(n,m)} \ .
\end{split}
\end{equation} 

Profile functions $K_1$ -- $K_4$ and $\Phi_1$, $\Phi_2$ 
depend on the coordinates $r$ and $\theta$ only. Thus, this ansatz is axially symmetric 
in a sence that a spacial rotation around the $z$-axis 
can be compencated by an Abelian gauge transformation
$U = \exp \{i\omega(r,\theta) \tau_\vphi^{(n)}/2\}$ which leaves the ansatz form-invariant.   

To obtain a regular solution we make use of the $U(1)$ gauge symmetry.
to fix the gauge \cite{KKT}.  We impose the condition 
$$
G_f = \frac{1}{r^2}\left(r\partial_r K_1 - \partial_\theta K_2\right) = 0 \ .
$$
The gauge fixing term $L_\eta = \eta G_f^2$ must be added to the Lagrangian (\ref{LagrGG}). 

With this Ansatz the field strength tensor components become
\begin{equation}
\begin{split}
F_{r\theta} &= -\frac{1}{r}\left(\partial_\theta K_1 - r \partial_r K_2\right)\frac{\tau_\vphi^{(n)}}{2}\\
F_{r\vphi} &= -\frac{n}{r}\sin\theta\biggl[\left(K_1 \frac{\sin(m\theta)}{\sin\theta} + K_1(K_4-1)
            -  r \partial_r K_3\right)\frac{\tau_r^{(n,m)}}{2}\\
           &+\left(K_1 \frac{\cos(m\theta)}{\sin\theta} + K_1K_3 + r \partial_r K_4\right)
\frac{\tau_\theta^{(n,m)}}{2}\biggr];\\
F_{\theta\vphi} &= n \sin\theta\biggl[\left((1-K_2)\frac{\sin(m\theta)}{\sin\theta} + 
                 (1-K_4)(K_2+n-1)-\partial_\theta K_3 - K_3 \cot\theta\right)\frac{\tau_r^{(n,m)}}{2}\\
                & + \left((1-K_2)\frac{\cos(m\theta)}{\sin\theta} - K_3(K_2+n-1)+ \partial_\theta K_4 
                    -  (1-K_4)\cot\theta\right)
\frac{\tau_\theta^{(n,m)}}{2}\biggr] \ .
\end{split}
\end{equation}
and the components of covariant derivative of the Higgs field
\begin{equation}
\begin{split}
D_r\Phi &= \frac{1}{r}\left(\left[r \partial_r \Phi_1 + K_1\Phi_2\right] \tau_r^{(n,m)} + 
        \left[r \partial_r \Phi_2 - K_1\Phi_1\right]\tau_\theta^{(n,m)}\right);\\
D_\theta \Phi &= \left[\partial_\theta \Phi_1 - K_2\Phi_2\right]\tau_r^{(n,m)} + 
                 \left[ \partial_\theta \Phi_2 + K_2\Phi_1\right]\tau_\theta^{(n,m)};\\
D_\vphi \Phi &=m\left[\sin\theta(K_3\Phi_2 + K_4\Phi_1) + \cos \theta \Phi_2\right]\tau_\vphi^{(n)} \ .
\end{split}
\end{equation}

Variation of the Lagrangian (\ref{LagrGG}) with respect to the profile functions yelds a  
system of six second order non-linear partial differential equations in the coordinates $r$ and $\theta$.

\subsubsection{Boundary conditions}
To obtain regular solutions with finite energy density and correct asymptotic 
behavour we impose the boundary conditions. 
Regularity at the origin requires
$$
K_1(0,\theta)=0\ , \ \ \ \ K_2(0,\theta)= 1 \ , \ \ \ \
K_3(0,\theta)=0 \ , \ \ \ \ K_4(0,\theta)=1 \ , \ \ \ \
$$
$$
\sin(m\theta) \Phi_1(0,\theta) + \cos(m\theta) \Phi_2(0,\theta) = 0 \ ,
$$
$$
\left.\partial_r\left[\cos(m\theta) \Phi_1(r,\theta)
              - \sin(m\theta) \Phi_2(r,\theta)\right] \right|_{r=0} = 0 
$$
that is $\Phi_\rho(0,\theta) =0$. 

To obtain the boundary conditions at infinity we require that
solutions in the vacuum sector ($m=2k$) tend to
a gauge transformed trivial solution, 
$$
\Phi \ \longrightarrow U \tau_z U^\dagger \   , \ \ \
A_\mu \ \longrightarrow  \ i \partial_\mu U U^\dagger \ ,
$$
and the solutions in the topological charge $n$ sector ($m=2k+1$)
to tend to
$$
\Phi  \longrightarrow  U \Phi_\infty^{(1,n)} U^\dagger \   , \ \ \
A_\mu \ \longrightarrow \ U A_{\mu \infty}^{(1,n)} U^\dagger
+i \partial_\mu U U^\dagger \  ,
$$
where
$$ \Phi_\infty^{(1,n)} =\tau_r^{(1,n)}\ , \ \ \
A_{\mu \infty}^{(1,n)}dx^\mu =
\frac{\tau_\vphi^{(n)}}{2} d\theta
- n\sin\theta \frac{\tau_\theta^{(1,n)}}{2} d\vphi
$$
is the asymptotic solution of a charge $n$ multimonopole,
and $SU(2)$ matrix  $U = \exp\{-i k \theta\tau_\vphi^{(n)}\}$, both
for even and odd $m$.
Consequently, solutions with even $m$ have vanishing magnetic charge,
whereas solutions with odd $m$ possess magnetic charge $n$.

In terms of the functions $K_1 - K_4$, $\Phi_1$, $\Phi_2$ these boundary
conditions read

\begin{equation}
K_1 \longrightarrow 0 \ , \ \ \ \
K_2 \longrightarrow 1 - m \ , \ \ \ \
\label{K12infty}
\end{equation}
\begin{equation}
K_3 \longrightarrow \frac{\cos\theta - \cos(m\theta)}{\sin\theta}
\ \ \ m \ {\rm odd} \ , \ \ \
K_3 \longrightarrow \frac{1 - \cos(m\theta)}{\sin\theta}
\ \ \ m \ {\rm even} \ , \ \ \
\label{K3infty}
\end{equation}
\begin{equation}
K_4 \longrightarrow 1- \frac{\sin(m\theta)}{\sin\theta} \ ,
\label{K4infty}
\end{equation}
\begin{equation}        \label{Phiinfty}
\Phi_1\longrightarrow  1 \ , \ \ \ \ \Phi_2 \longrightarrow 0 \ .
\end{equation}

Regularity on the $z$-axis, finally, requires
$$
K_1 = K_3 = \Phi_2 =0 \ , \ \ \  \
\partial_\theta K_2 = \partial_\theta K_4 = \partial_\theta \Phi_1 =0 \ ,
$$
for $\theta = 0$ and $\theta = \pi$.

Subject of the above boundary conditions, we
constructed numerically solutions with $1\leq m \leq 6$, $1 \leq n \leq 6$  
and several values of the Higgs selfcoupling constant $\lambda$. 

\subsubsection{Properties of the solutions}
We give here a description of the general properties of the solutions.
Note that asymptotic behavior of the profile functions allows to 
check the equivalence between the topological charge $Q$  
and magnetic charge $g$. Indeed,
\begin{equation}\label{firstcharge}
\begin{split}
Q &= \frac{1}{8\pi}\int\limits_{S_2} d^2\xi~ \varepsilon_{\alpha\beta}\varepsilon_{abc}
\phi^a\partial_\alpha\phi^b\partial_\beta \phi^c = \\
  & = \frac{2nm}{8\pi}\int d\theta d \vphi~[H_1\sin(m\theta)+H_2\cos(m\theta)]=
\frac{n}{2} \left[1-(-1)^m\right];\\
g &= \frac{1}{4\pi }\int 
\Tr \,\varepsilon_{ijk} \left( F_{ij} D_k \hat \Phi \right) d^3 r  = 
\frac{1}{2\pi }\int \limits_{S_2} d\theta d\vphi ~\Tr(F_{\theta\vphi}\Phi)\\
   &=\frac{n}{2}\int d\theta \left(\sin(m\theta) - \partial_\theta(\sin\theta K_3)\right) 
= \frac{n}{2}\left[1 - (-1)^m\right] 
\end{split}
\end{equation}
where we used the definitions (\ref{g-deff}), (\ref{charge-g}) and substitute the
asymptotic behavior of the provile functions (\ref{K3infty}),(\ref{K4infty})  
(\ref{Phiinfty}). Thus, the configurations given by the axially symmetric 
Ansatz (\ref{ansatzA}),(\ref{ansatzPhi})
are either deformations
of the topologically trivial sector (e.g. monopole-antimonopole pair, $n=1$, $m=2$)
or deformations of core of charge $n$ Rebbi-Rossi multimonopoles (e.g. separation of two zeros 
of the Higgs field of the charge 2 monopole).

\subsubsection{2d Poincare index of the vector field}
The axial symmetry of the ansatz (\ref{ansatzA}),(\ref{ansatzPhi}) means that we can choose 
any value of the azimuthal angle, for example $\vphi = 0$, that is consider 
the behavour of the fields on the $xz$ plane.   
Then classification of the solutions can be constructed 
by making use of the {\it 2d Poincare index} of the Higgs field on that plane which supplements 
3d topological characteristic (\ref{3dindex}) \cite{Milnor}.

By analogy with (\ref{3dindex}),  2d index is defined  by 
consideration of a smooth two-dimensional  vector field ${\vec v}(x) = (v_1, v_2)$ 
on a compact space $X$ with an isolated zero at 
$x_0$. The function $f(x) = \frac{{\vec v}(x)}{|v(x)|}$ maps a small circle  $S^1$ 
centered at $x_0$ into the 
unit sphere $S^1_\phi$. The degree of this mapping is called the Poincare index $i_{x_0}$ of vector field 
at the zero  $x_0$. If the circle is parametrized by an angle $\alpha \in [0:2\pi]$, the index is 
\begin{equation}
\label{index}
i_{x_0} = \frac{1}{2\pi}\int\limits_{S_1}\frac{{\vec v}\wedge d {\vec v}}{|v|^2} = 
\frac{1}{2\pi}\int d\alpha~ \frac{v_1 d_\alpha v_2 - v_2 d_\alpha v_1}{v_1^2 + v_2^2}
\end{equation}
Note that different orientation of the vector field at the point $\alpha =0$ on 
the small circle $S_1$ yields configurations which are labeled by the same index 
(see Fig.~\ref{f-1}).  Any perturbation of the contour without
crossing a node of the 2-dimensional vector field does not change the index. 

\begin{figure}[t]
\begin{center}
\setlength{\unitlength}{1cm}
\lbfig{f-1}
\begin{picture}(13,4.2)
\put(1.7,0.0)
{\mbox{
\psfig{figure=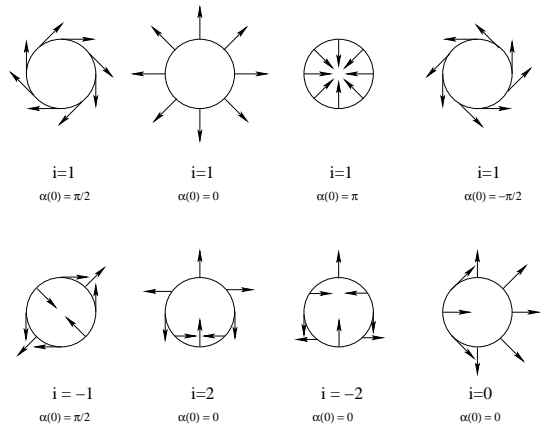,height=5.6cm}}}
\end{picture}
\caption{
Indices of two dimensional vector field.}
\end{center}
\end{figure}
Let us consider a closed contour near spacial boundary of a 
plane which encircles some set of isolated zeros  
of a two-dimensional vector field  \footnote{This corresponds to an infinitesimal contour 
around the pole of sphere $S_2$ which can be obtained by compactification of that plane.  
Then the vector field on $S_2$ is pointed outward at all boundary points and the sum of all indices 
of zeros of such a vector field is equal to the Euler number $\chi(S_2)=2$, a topological invariant
of $X$, which does not depend on particular choice of vector field. 
This is the original statement of the Poincare-Hopf theorem.}  

Mapping of a vector field ${\vec v}(x)$ at the spacial infinity 
into a unit circle yields the index $i_\infty$.  This index  will not change upon a
perturbation of the field, as long as no node crosses the contour during
perturbation. The Poincare-Hopf theorem then can be formulated
as a statement that the index of that field with respect to a contour near infinity 
is equal to the sum of the indices computed locally around each node of the field:
$$
i_\infty = \sum\limits_{k} i(x_0^{(k)})
$$

In the case under consideration we consider nodes of the Higgs field on the $xz$ plane. 
The boundary condition (\ref{Phiinfty}) means that $\Phi(r,\theta) \to \tau_r^{(n,m)}$ 
as $r \to \infty$. Thus, $\Phi_y = 0$ and coresponding 
components of two-dimensional vector field are 
$v_1 \equiv \Phi_z = \cos (m\theta),v_2 \equiv  \Phi_x = \sin(m\theta)$ as $x > 0$ and 
$v_1 \equiv \Phi_z = \cos (m\theta), v_2 \equiv  \Phi_x = (-1)^n \sin(m\theta)$ 
as $x < 0$ respectively. 
Then calculation of the index 
$i_\infty$ according to eq. (\ref{index}) yields
\begin{equation} \label{secondcharge}
i_\infty = \frac{m}{2}\left[1 -  (-1)^n\right]
\end{equation}
This is a 2-dimensional topological characteristics of the axially symmetric 
solutions which is complementary to the 
3-dimensional invariant (\ref{3dindex}). Note that it is ``dual'' to the 
topological charge (\ref{firstcharge}) in a sence that permutation $n \rightleftharpoons m$ 
yields $i_\infty \rightleftharpoons Q$. However, the 
index $i_\infty $ is not an invariant of the transformations between homotopically 
related configurations within 
the same topological sector. Corresponding deformations of 3-dimensional vector field may be related with  
vanishing of the projection of the field on $xz$ plane on the circle at the spacial infinity. The crossing 
of this node yields jump of the 2-dimensional index $i_\infty $  by some integer value.

\subsubsection{$m$-chains}

The solution of the first type are $m$-chains constructed in \cite{KKS1,KKS2}.
They are characterized by $\theta$ winding number $m>1$ 
and $\vphi$ winding number $n=1,2$. 
 
Let us consider $n=1$ configurations first. 
These $m$-chains possess $m$ nodes of the Higgs field on the $z$-axis.
Due to reflection symmetry, each node on the negative $z$-axis corresponds
to a node on the positive $z$-axis.
The nodes of the Higgs field $x_0{(k)}$ are associated with the location of the
monopoles and antimonopoles and all of them are characterized by the index  $i_(x_0^{(k)}) = 1$ 
(cf Fig. 2).  
The index $i_\infty$ of these $m$-chains is obviously equal to the winding number $m$ whereas 
the topological charge is either unity (for odd $m$) or zero (for even $m$).  

Indeed, for odd $m$ ($m=2k+1$) the Higgs field possesses 
$k$ nodes on the positive $z$-axis and one node at the origin.
The node at the origin corresponds
to a monopole if $k$ is even and to an antimonopole if $k$ is odd.
For even $m$ ($m=2k$) there is no node of the Higgs field at the origin. 
 
The $m=1$ solution is the spherically symmetric 't Hooft-Polyakov monopole
which we discussed in the first part of our review.
The $m=3$ (M-A-M) and $m=5$ (M-A-M-A-M) chains represent
saddlepoints with unit topological charge. 
The $m=2$ chain is identical to the 
monopole-antimonopole (M-A) pair discussed in \cite{Rueber85,mapKK}.
The M-A pair as well as the $m=4$ (M-A-M-A) and $m=6$ (M-A-M-A-M-A)
chains form saddlepoints in the vacuum sector.

In Fig. \ref{f-3}~  we present the dimensionless energy density and nodes of the Higgs field
for the $n=1$ solutions with $\theta$ winding number $m=1, \dots , 6$. 
The energy density of the $m$-chain possesses $m$ maxima on the $z$-axis,
and decreases with increasing $\rho$.
The locations of the maxima are close to the nodes of the Higgs field.
For a given $m$ the maxima are of 
similar magnitude, but their height decreases with increasing $m$. Increasing 
of $\lambda$ makes these maxima sharper and decreases the distance between 
the locations of the monopoles. We observe that for a given $\lambda$ 
the distances between the corresponding nodes increase with increasing $m$.

\begin{figure}[t]
\begin{center}
\setlength{\unitlength}{1cm}
\lbfig{f-3}
\begin{picture}(0,3.0)
\put(-7.5,0.0)
{\mbox{\epsfysize=4.0cm\epsffile{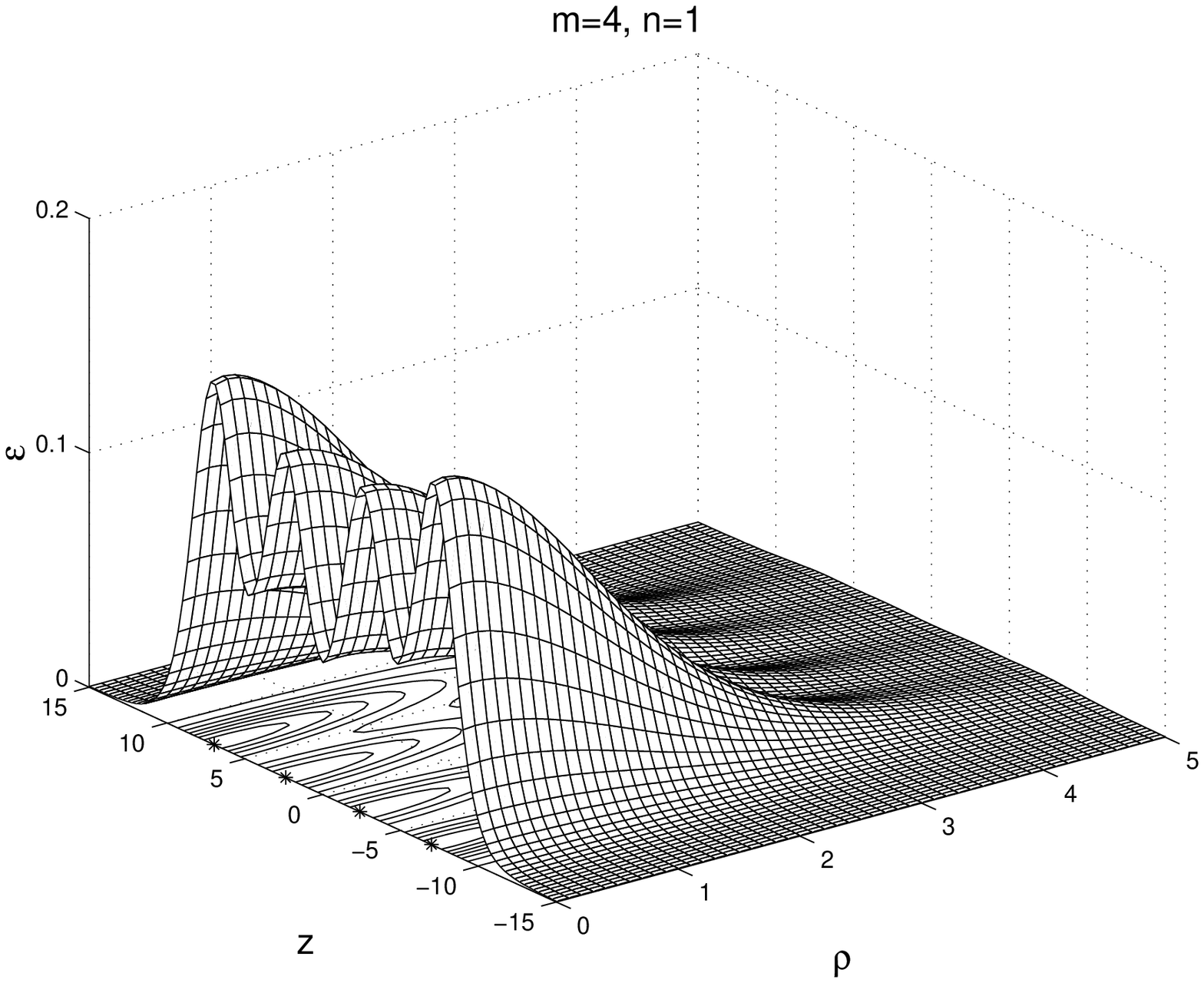}}}
\end{picture}
\begin{picture}(0,0.0)
\put(-2.5,0.0)
{\mbox{\epsfysize=4.0cm\epsffile{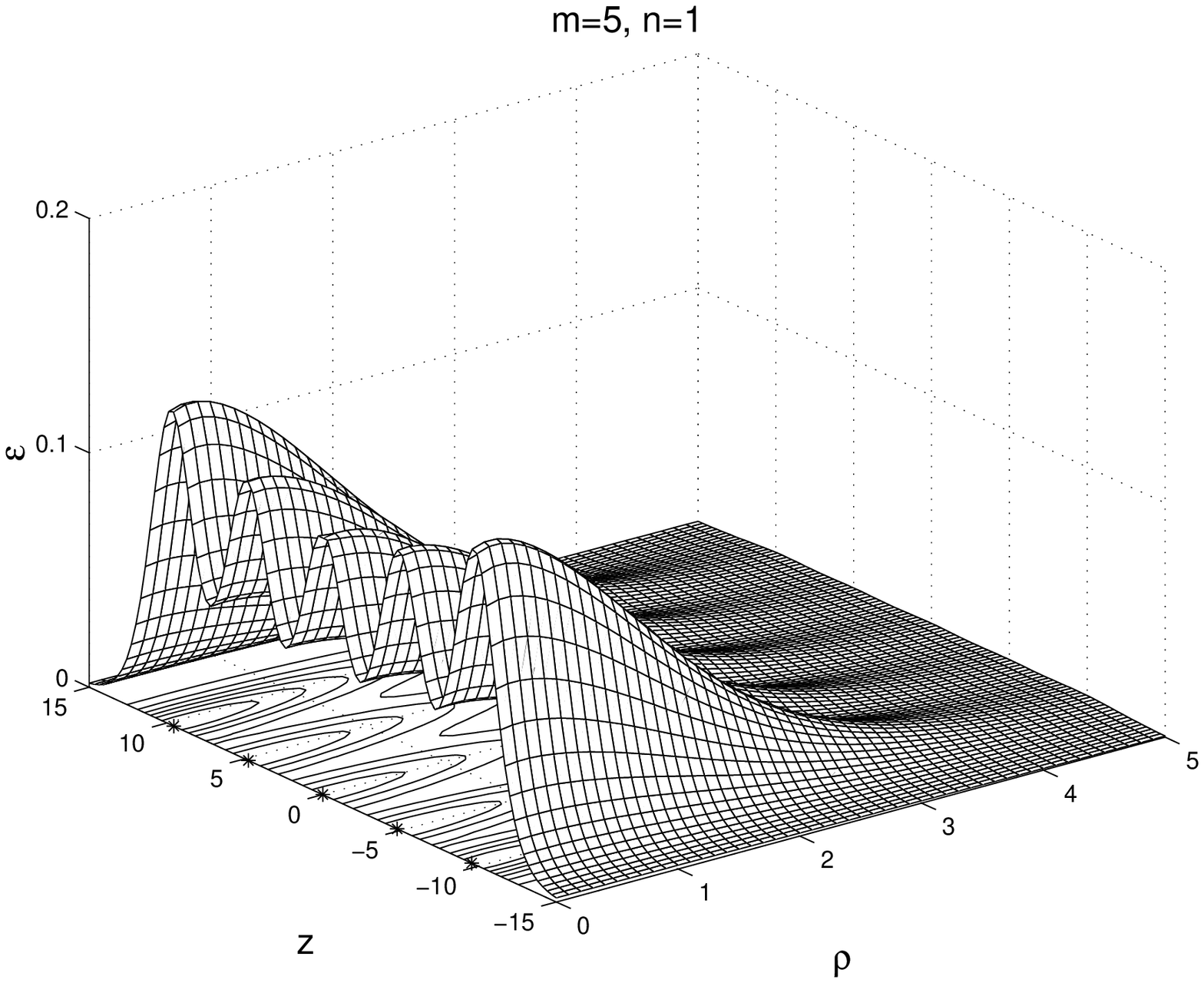}}}
\end{picture}
\begin{picture}(0,7.0)
\put(2.5,0.0)
{\mbox{\epsfysize=4.0cm\epsffile{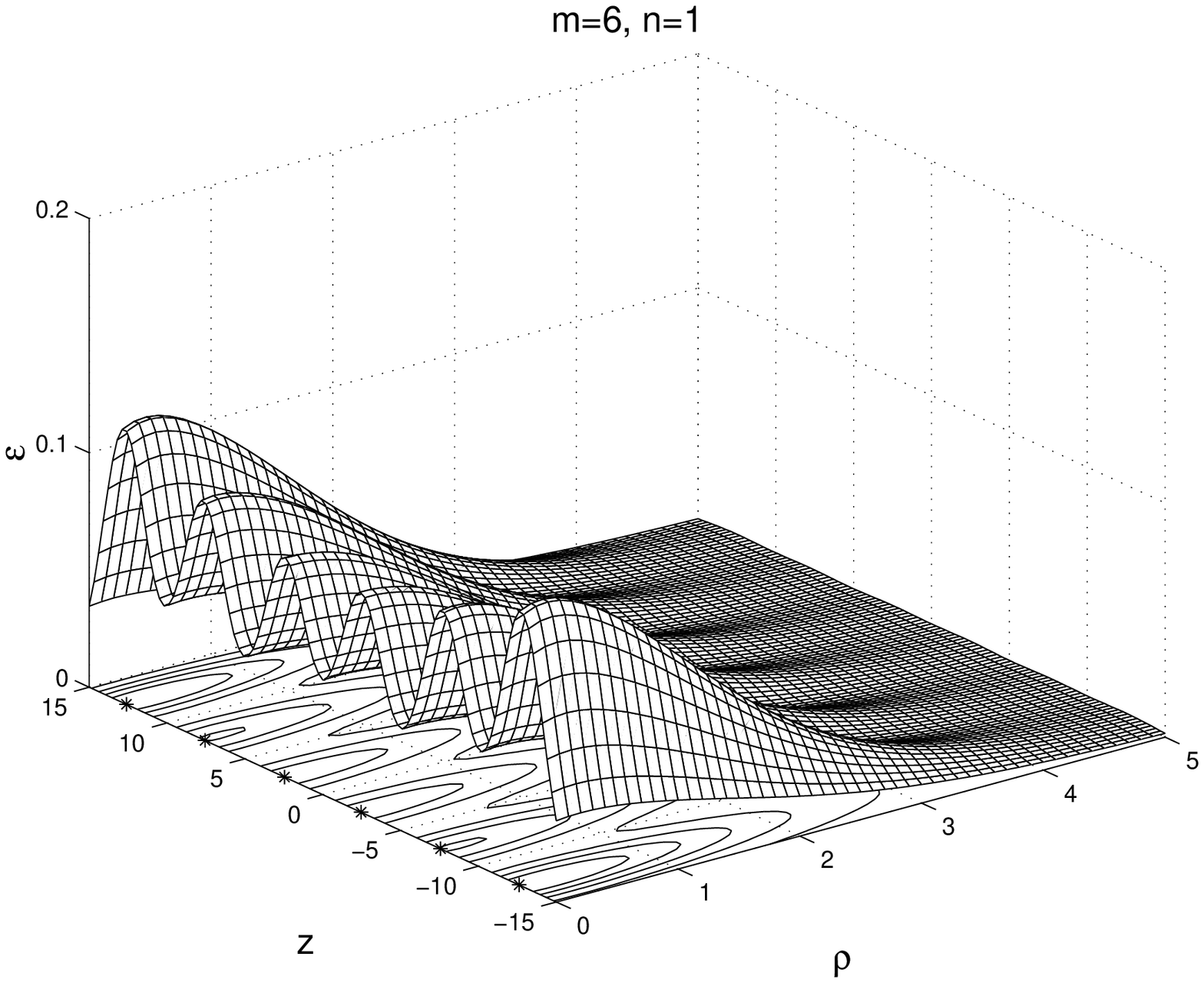}}}
\end{picture}
\begin{picture}(0,7.0)
\put(-7.7,5.0)
{\mbox{\epsfysize=4.0cm\epsffile{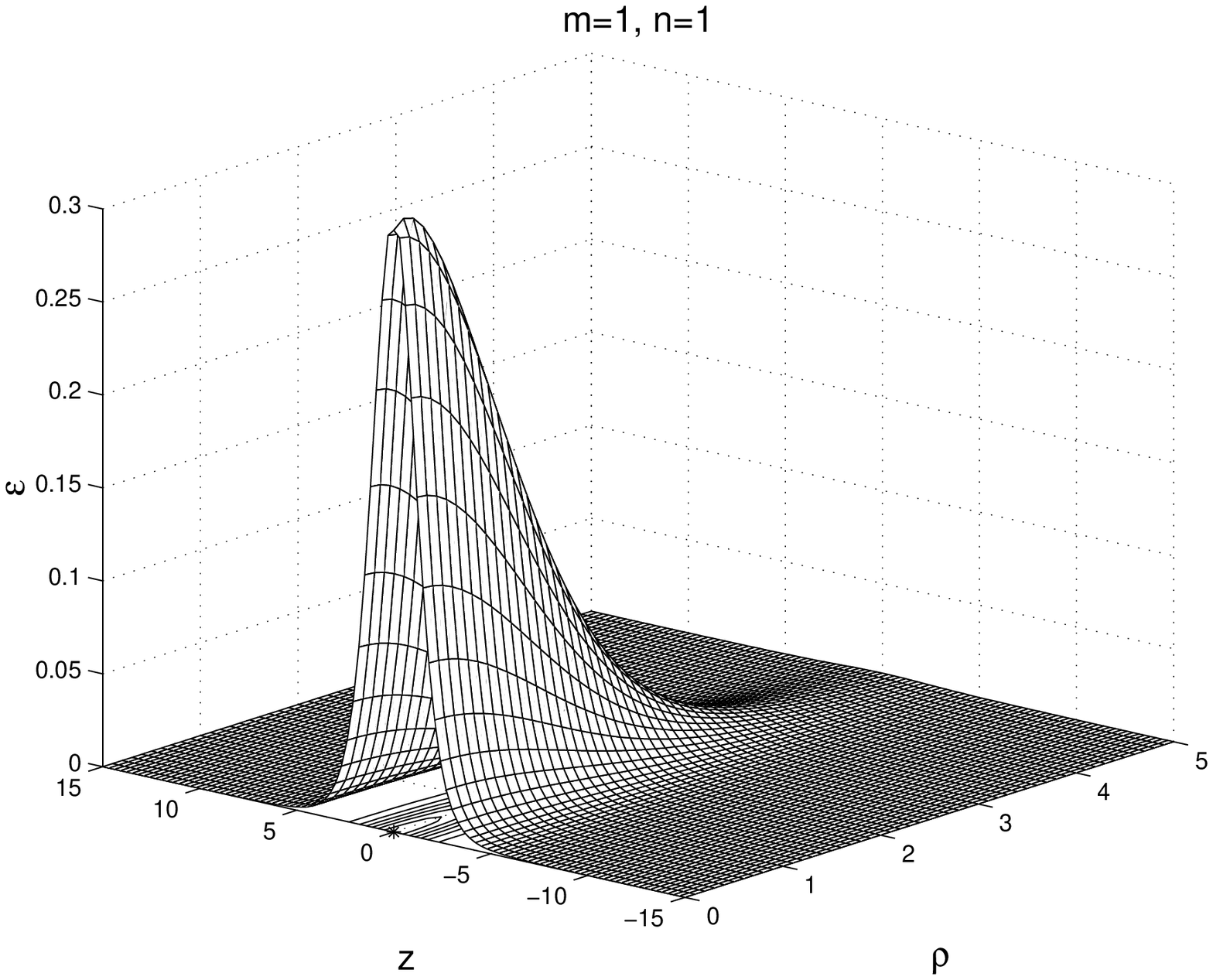}}}
\end{picture}
\begin{picture}(0,0.0)
\put(-2.7,5.0)
{\mbox{\epsfysize=4.0cm\epsffile{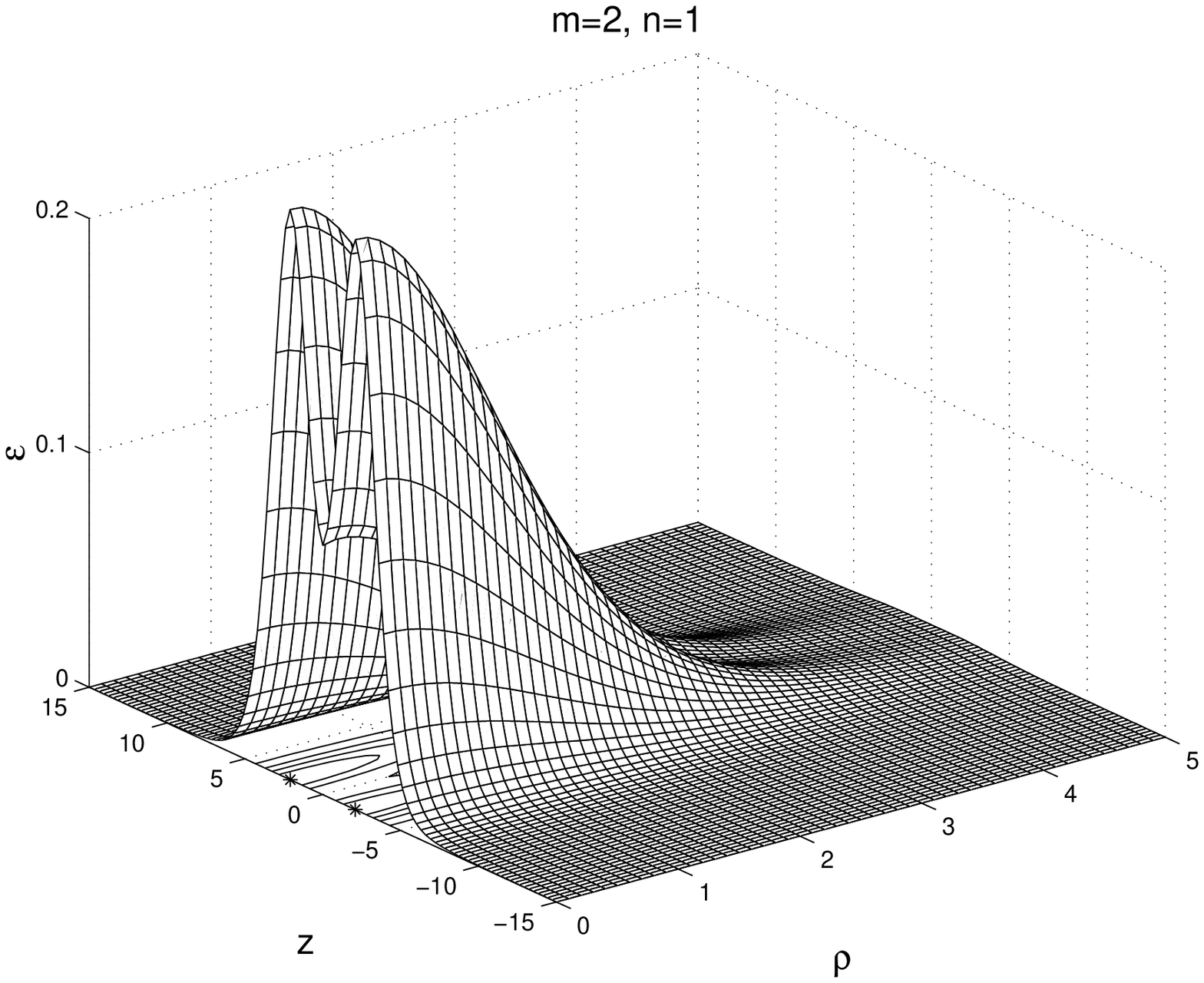}}}
\end{picture}
\begin{picture}(0,7.0)
\put(2.3,5.0)
{\mbox{\epsfysize=4.0cm\epsffile{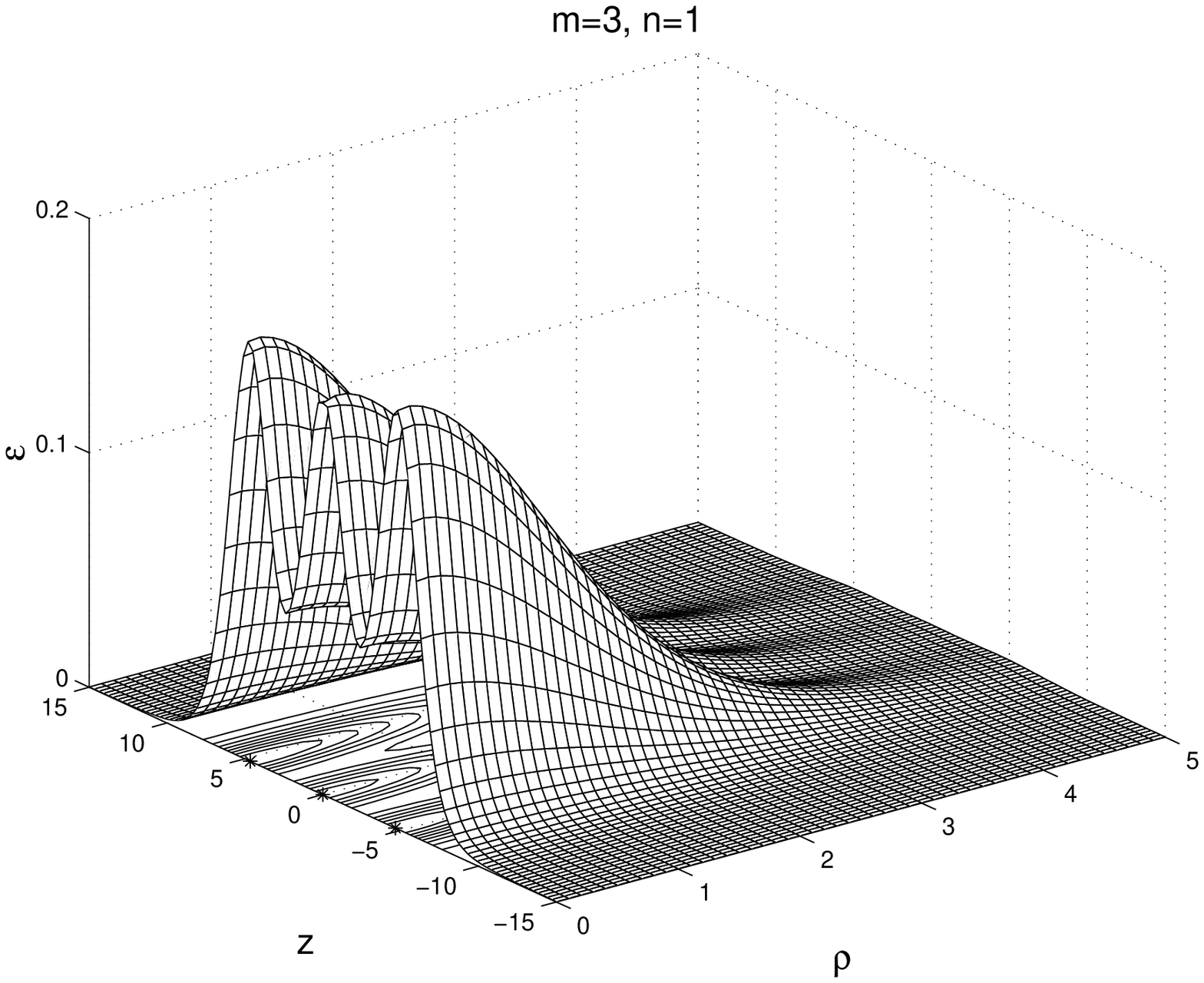}}}
\end{picture}
\caption{The rescaled energy density $E(\rho,z)$ is shown for the monopole-antimonopole chains
 with winding numbers $n=1,m=1,\dots 6$ at $\lambda = 0$.
}
\end{center}
\end{figure}

We observe that the energy  $E^{(m)}$ of an $m$-chain
is always smaller than the energy 
of $m$ single monopoles or antimonopoles (with infinite separation
between them),
i.~e.~$E^{(m)} < E_\infty = 4\pi\eta m$. 
On the other hand $E^{(m)}$ exceeds the minimal energy bound
given by the Bogolmol'ny limit $E_{\rm min}/4\pi\eta = 0$ for even $m$,
and $E_{\rm min}/4\pi\eta = 1$ for odd $m$.
We observe an (almost) linear dependence of the energy $E^{(m)}$ on $m$.
This can be modelled
by taking into account only the energy of $m$ single
(infinitely separated)
monopoles and the next-neighbour interaction between monopoles
and antimonopoles on the chain.
Defining the interaction energy as the binding energy of the 
monopole-antimonopole pair, 
$$
\Delta E = 2\ (4\pi \eta) - E^{(2)} \ ,
$$
we obtain as energy estimate for the $m$-chain
$$
E_{\rm est}^{(m)}/4\pi \eta  = m +(m-1) \Delta E \ .
$$ 
\begin{figure}[tbh]
\begin{center}
\setlength{\unitlength}{1cm}
\lbfig{f-4}
\begin{picture}(0,3.5)
\put(-6.5,-0.0)
{\mbox{
\psfig{figure=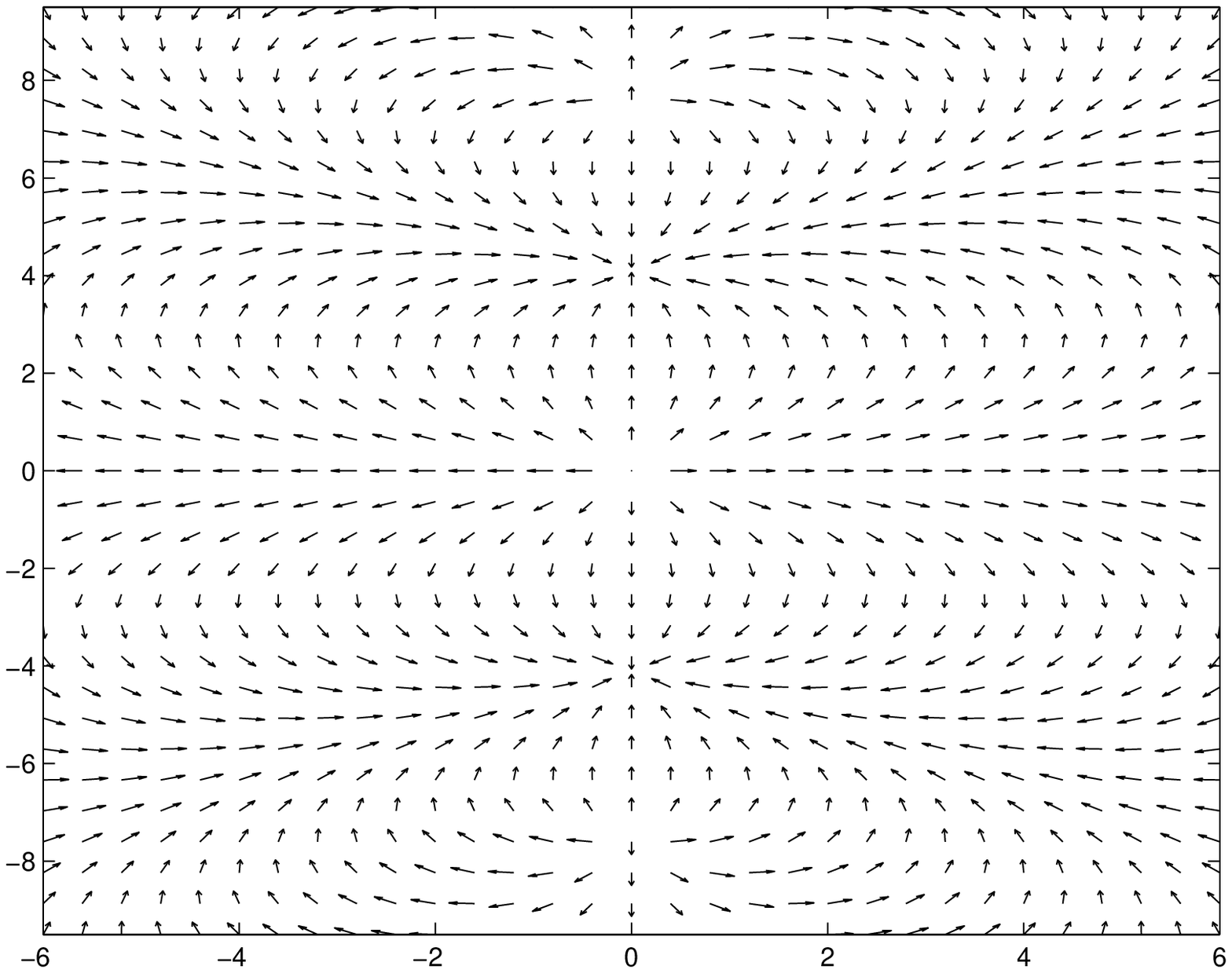,height=3.5cm, angle =0}}}
\end{picture}
\setlength{\unitlength}{1cm}
\begin{picture}(0,1.0)
\put(0.0,-0.0)
{\mbox{
\psfig{figure=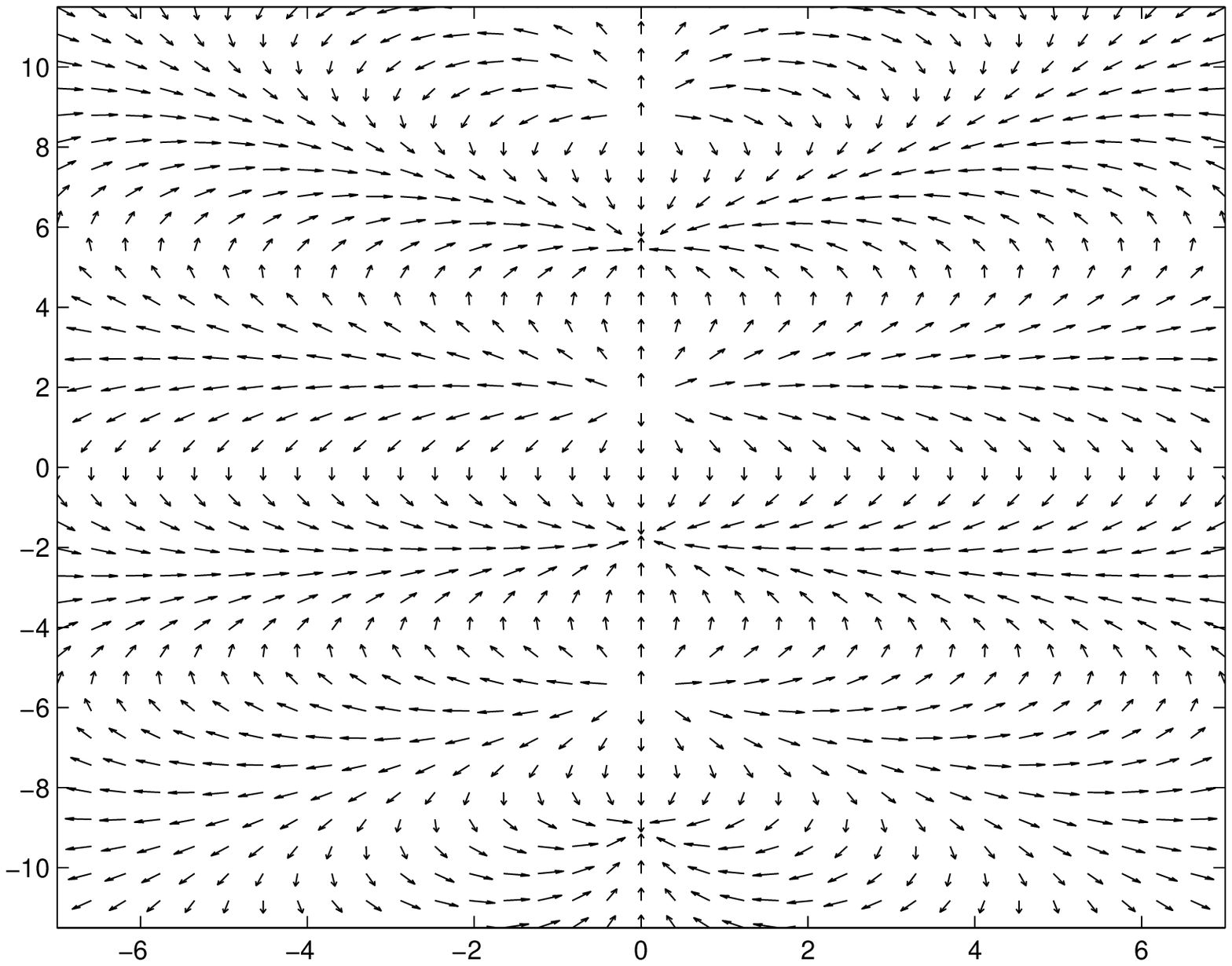,height=3.5cm, angle =0}}}
\end{picture}
\caption{2d Higgs field of monopole-antimonopole chains ($m=5,n=1$ and $m=6,n=1$)
All the nodes correspond to the unit index.
}
\end{center}
\end{figure}

We interpret the $m$-chains as equilibrium states of $m$ non-BPS
monopoles and antimonopoles. As shown long ago \cite{Manton77},
the force between BPS monopoles is given by twice the Coloumb force
when the charges are unequal, and vanishes when the charges are equal,
provided the monopoles well separated.  
Thus, monopoles and antimonopoles can only be 
in static equilibrium, if they are close enough to experience a
repulsive force that counteracts the attraction. Indeed, is the separation between 
monopoles is not too large, both photon and scalar particles remain massive,  
the vector bosons $A_\mu^\pm$ are ``defrosted'' and all the particles from the
spectrum contribute to the short-range Yukawa-type interactions.  

In other words, $m$-chains are essentially non-BPS solutions.
To see this in another way let us consider the limit $\lambda = 0$. Then 
the energy of these `non-Bogomol'nyi BPS monopoles' can be written in the form (\ref{E2}):
\begin{equation}
E  = \int\left\{ \frac{1}{4} \Tr\left(
\left(\varepsilon_{ijk} F_{ij}\pm D_k \Phi \right)^2 \right)
 \mp\frac{1}{2}\varepsilon_{ijk} \Tr\left( F_{ij} D_k \Phi \right)	
	\right\} d^3 r \ .
\label{E}
\end{equation}
The second term is proportional to the topological charge and 
vanishes when $m$ is even. The first term is just the integral
of the square of the Bogomol'nyi equations. Thus, for even $m$ 
the energy is a measure for the deviation of the solution from 
selfduality.

Let us now consider chains consisting of multimonopoles with winding number 
$n=2$. Identifying the locations of the Higgs zeros on the symmetry axis
with the locations of the monopoles and antimonopoles, we observe that
when each pole carries charge $n=2$,
the zeros form pairs, where the distance between the monopole and 
the antimonopole of a pair
is less than the distance to the neighboring monopole or antimonopole,
belonging to the next pair. 

We observe furthermore,
that the equilibrium distance of the monopole-antimonopole pair
composed of $n=2$ multimonopoles
is smaller than the equilibrium distance of the monopole-antimonopole pair 
composed of single monopoles.
Thus the higher attraction between the poles of a pair with 
charge $n=2$ is balanced by the repulsion only 
at a smaller equilibrium distance.

The difference from the chains composed of pairs of monopole-antimonopole 
with winding number $n=1$ we considered above, 
is that the index $i_\infty$ of the $m$-chains composed of $n=2$ multimonopoles is zero. 
Moreover, the index of each double node of the Higgs field, which is compatible 
with the symmetry conditions, is also zero. Another difference is that for $n=2$ chains
maxima of the energy density are no longer 
coincide with (double) zeros of the Higgs field. The latter still are placed 
on the $z$ axis (cf. Fig. 5).
\begin{figure}[t]
\begin{center}
\setlength{\unitlength}{1cm}
\lbfig{f-2}
\begin{picture}(0,3.9)
\put(-6.5,-0.0)
{\mbox{
\psfig{figure=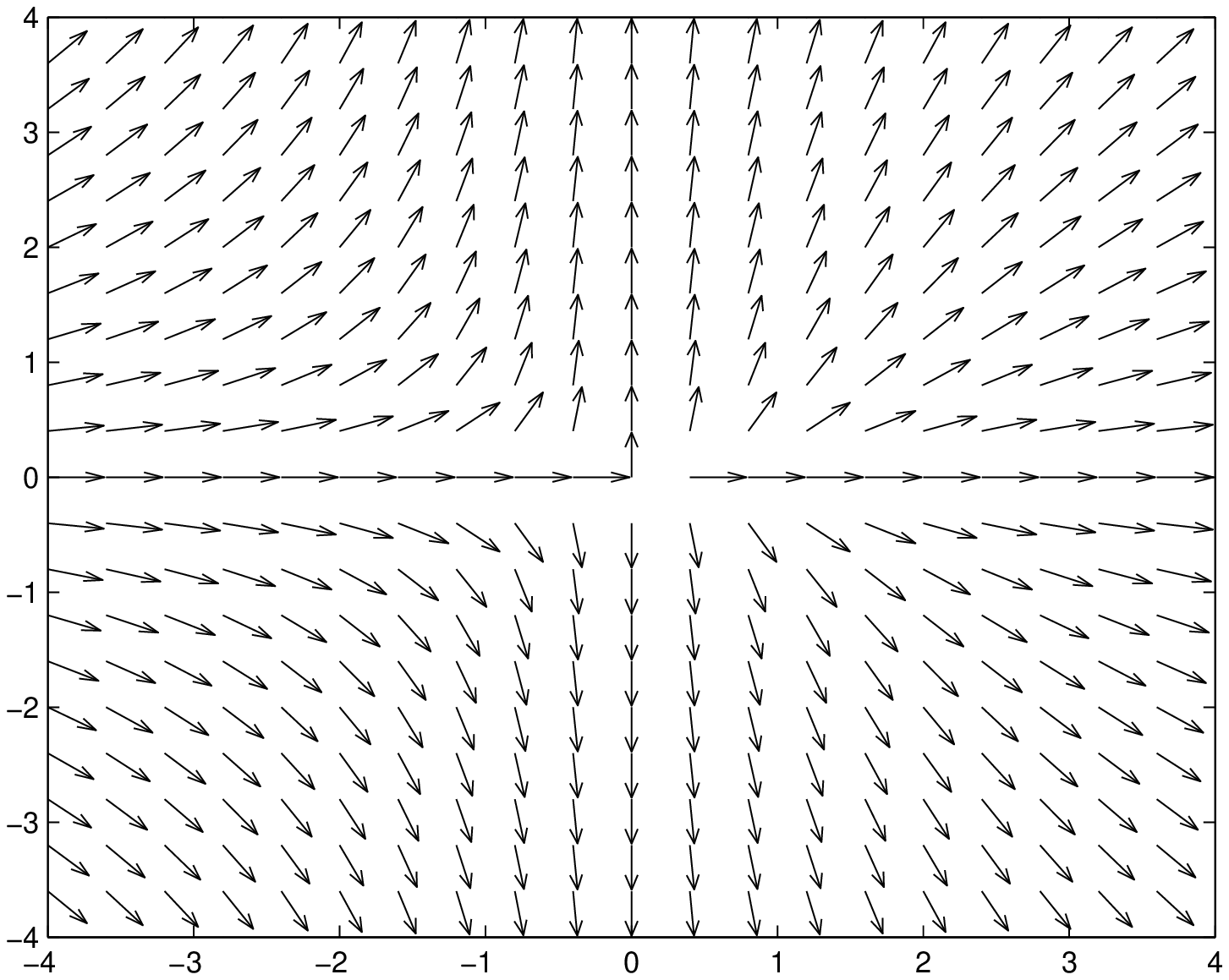,height=3.5cm, angle =0}}}
\end{picture}
\setlength{\unitlength}{1cm}
\begin{picture}(0,1.0)
\put(0.0,-0.0)
{\mbox{
\psfig{figure=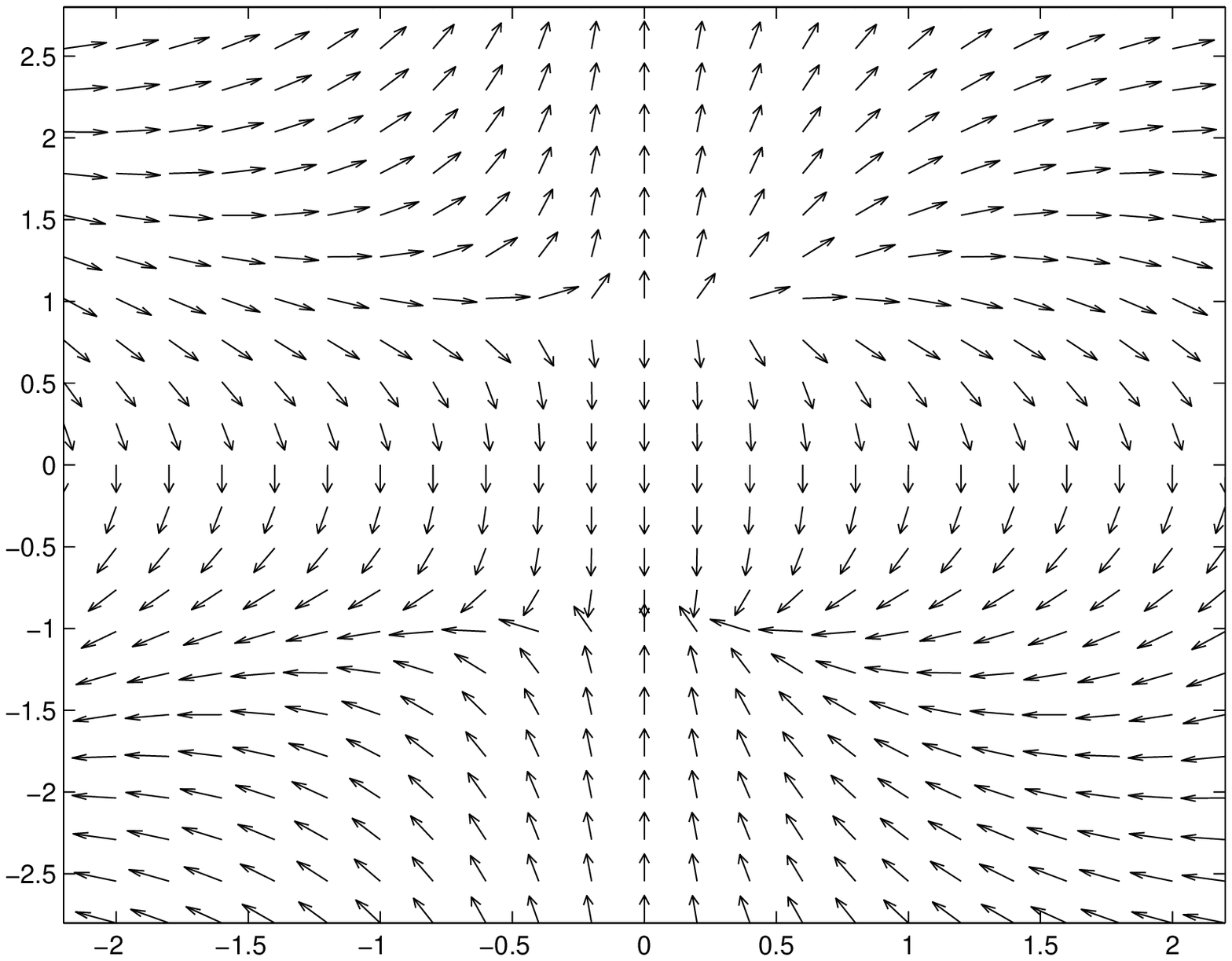,height=3.5cm, angle =0}}}
\end{picture}
\caption{2d Higgs field of a single 2-monopole and
2-monopole---2-antimonopole pair ($m=1,n=2$ and $m=2,n=2$)
The double nodes correspond to zero indices. 
}
\end{center}
\end{figure}

\subsubsection{Closed vortices}
The boundary conditions imposed at the origin on 
the Higgs field (\ref{ansatzPhi}) means that $\Phi_\rho(0,\theta) = 0$ and  
$$
\Phi (0,\theta) = \Phi_z (0,\theta) \tau_z^{(n,m)}
$$
Thus, the scalar field can either vanish there (for an odd $\theta$ winding number $m$), 
or to be directed along $z$ axis (for an even $m$).  In the former case there 
is a single $n$-monopole placed at the origin whereas the latter configuration 
in the limit of very large scalar coupling approaches the vacuum expectation value 
not only on the spacial boundary but also  in the vicinity 
of the origin. Then the solutions with different winding number $n$ link the trivial
configuration $\Phi(0,\theta) = (0,0, \Phi_z (0,\theta)$ and
its gauge rotated on the spacial infinity. For $n=1,2$ these solutions 
are monopole-antimonopole chains we discussed above. 
\begin{figure}[t]
\begin{center}
\setlength{\unitlength}{1cm}
\lbfig{f-04}
\begin{picture}(0,3.0)
\put(-7.5,3.7)
{\mbox{
\psfig{figure=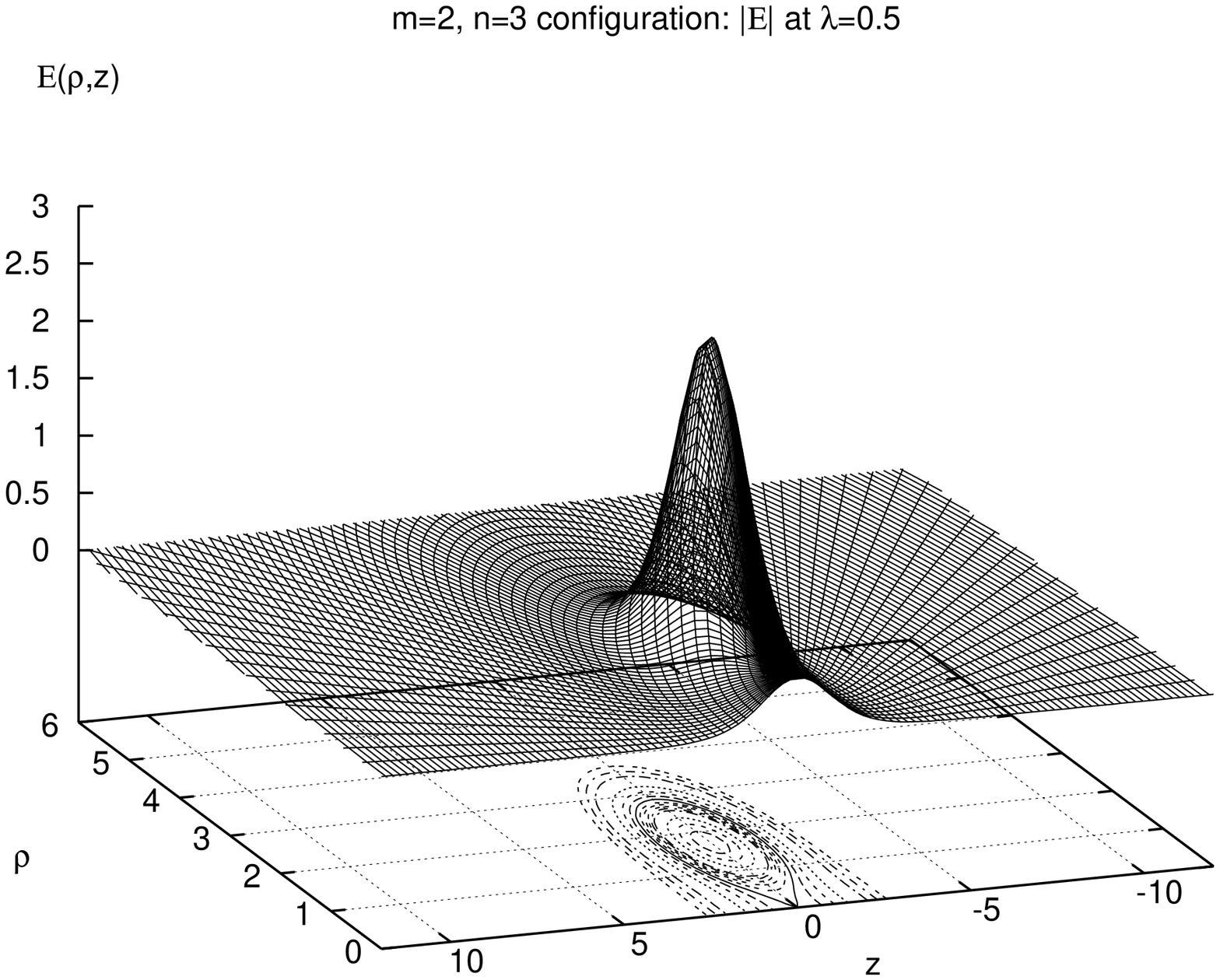,height=5.5cm, angle =-90}}}
\end{picture}
\begin{picture}(0,0.0)
\put(-2.5,3.7)
{\mbox{
\psfig{figure=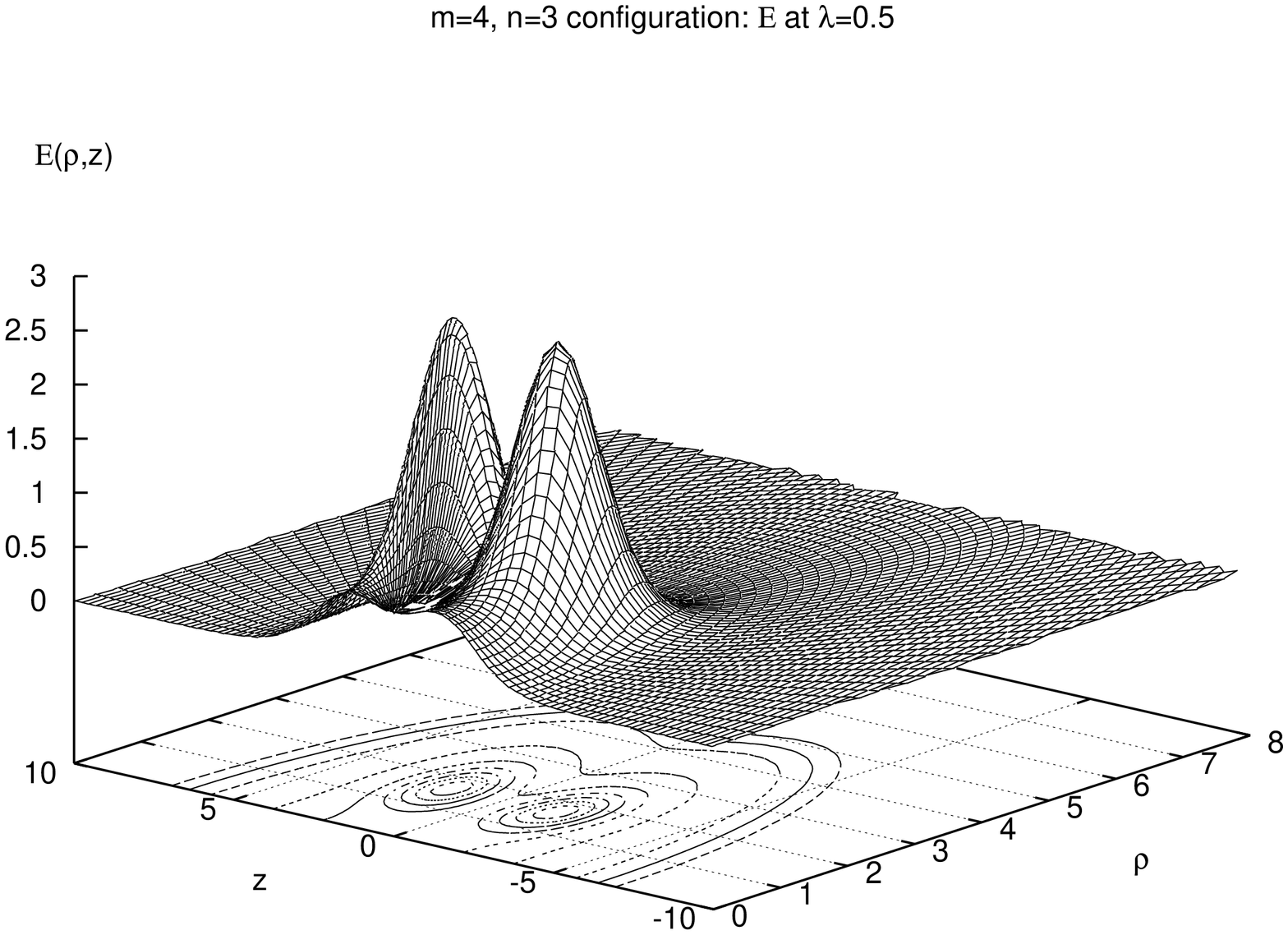,height=5.5cm, angle =-90}}}
\end{picture}
\begin{picture}(0,7.0)
\put(2.5,3.7)
{\mbox{
\psfig{figure=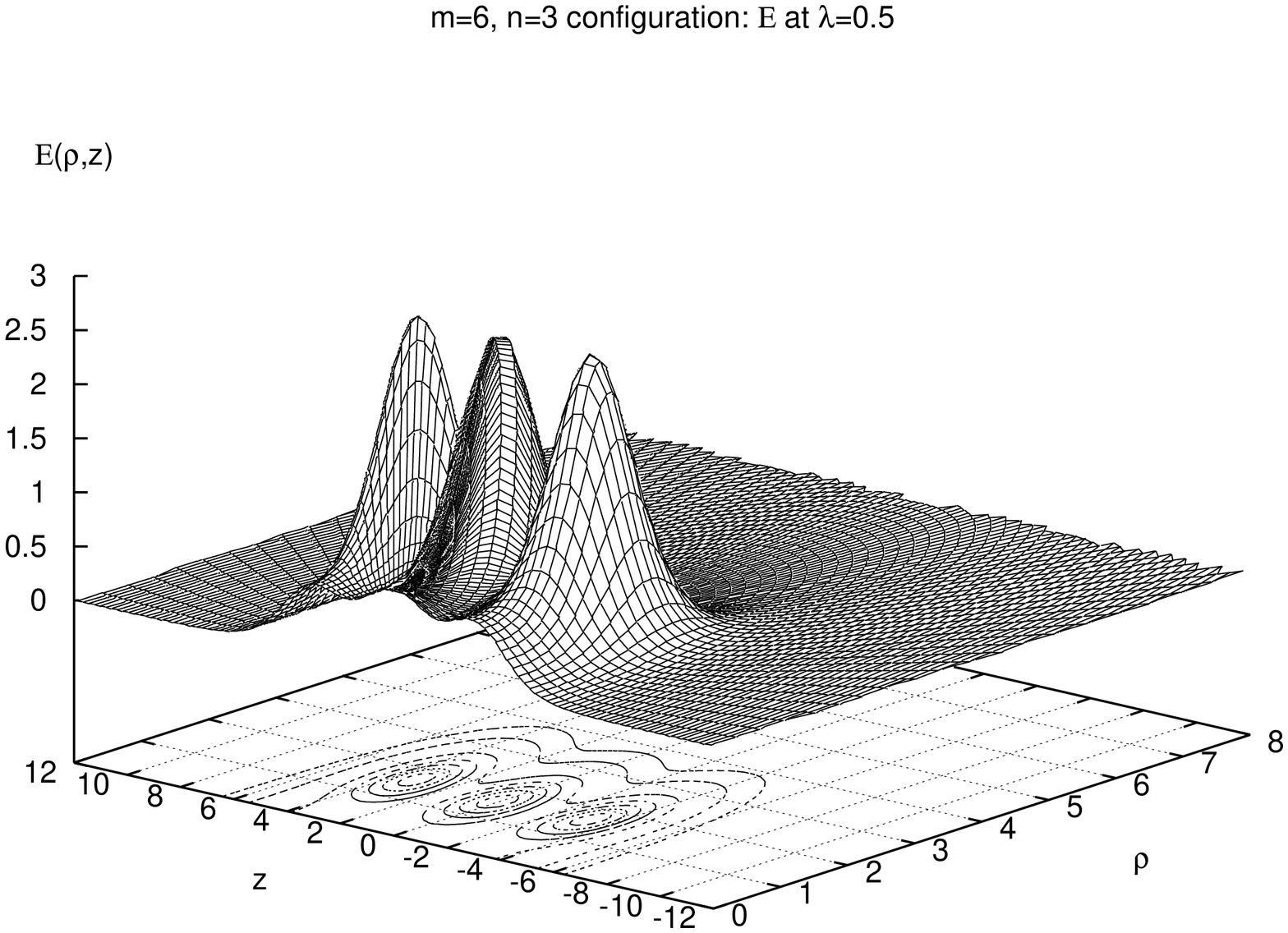,height=5.5cm, angle =-90}}}
\end{picture}
\begin{picture}(0,7.0)
\put(-7.7,7.5)
{\mbox{
\psfig{figure=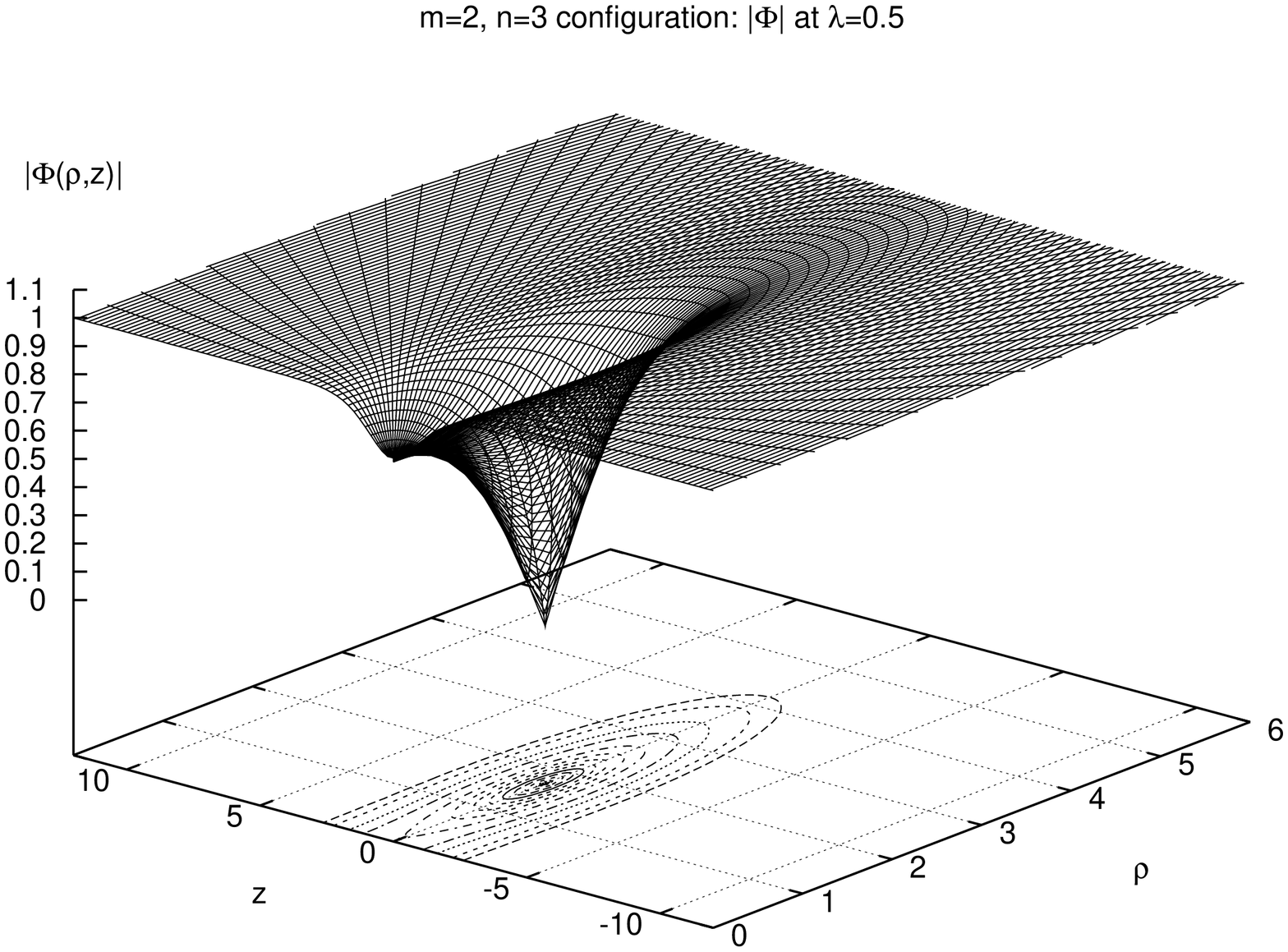,height=5.5cm, angle =-90}}}
\end{picture}
\begin{picture}(0,0.0)
\put(-2.7,7.5)
{\mbox{
\psfig{figure=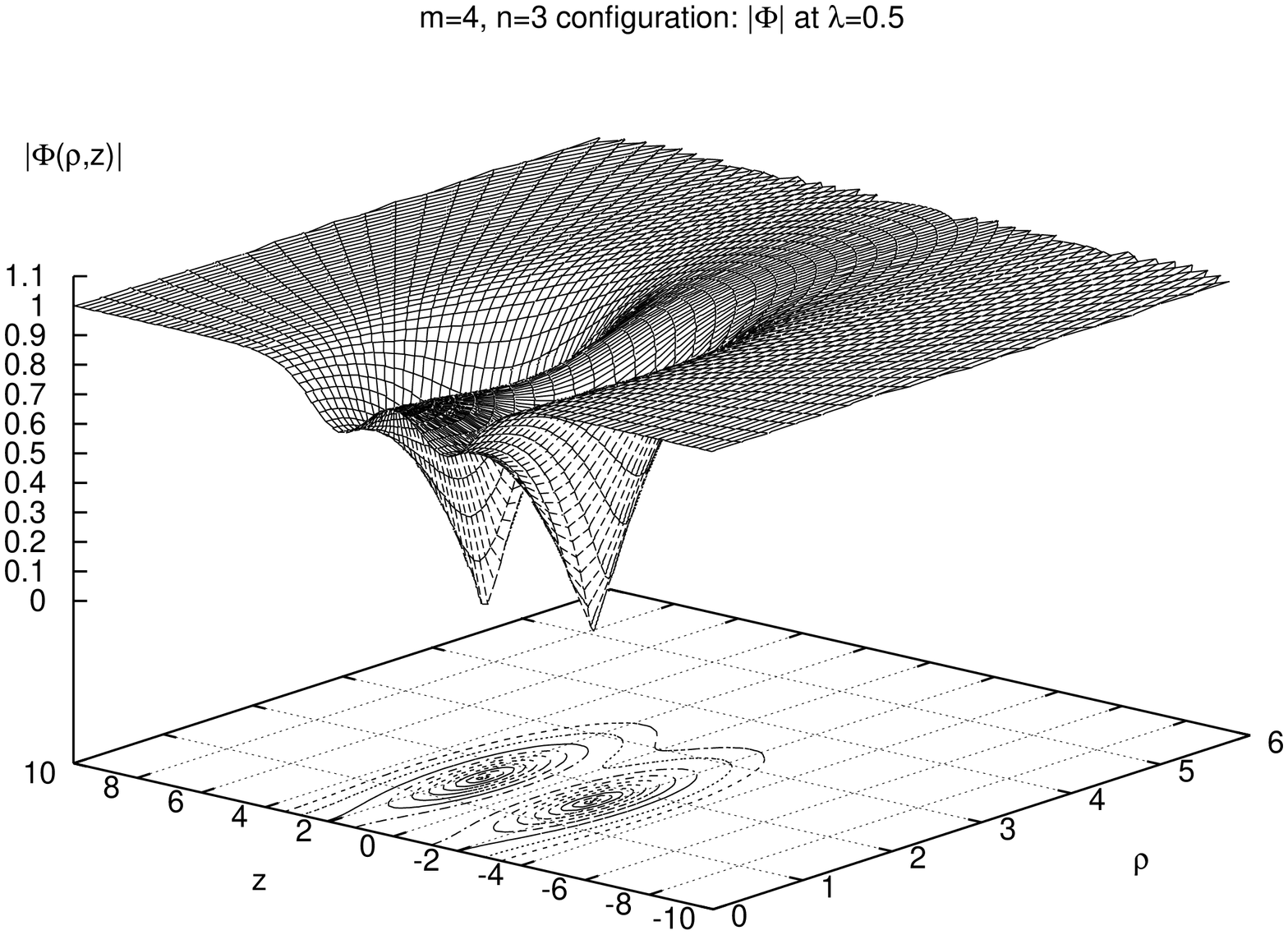,height=5.5cm, angle =-90}}}
\end{picture}
\begin{picture}(0,7.0)
\put(2.3,7.5)
{\mbox{
\psfig{figure=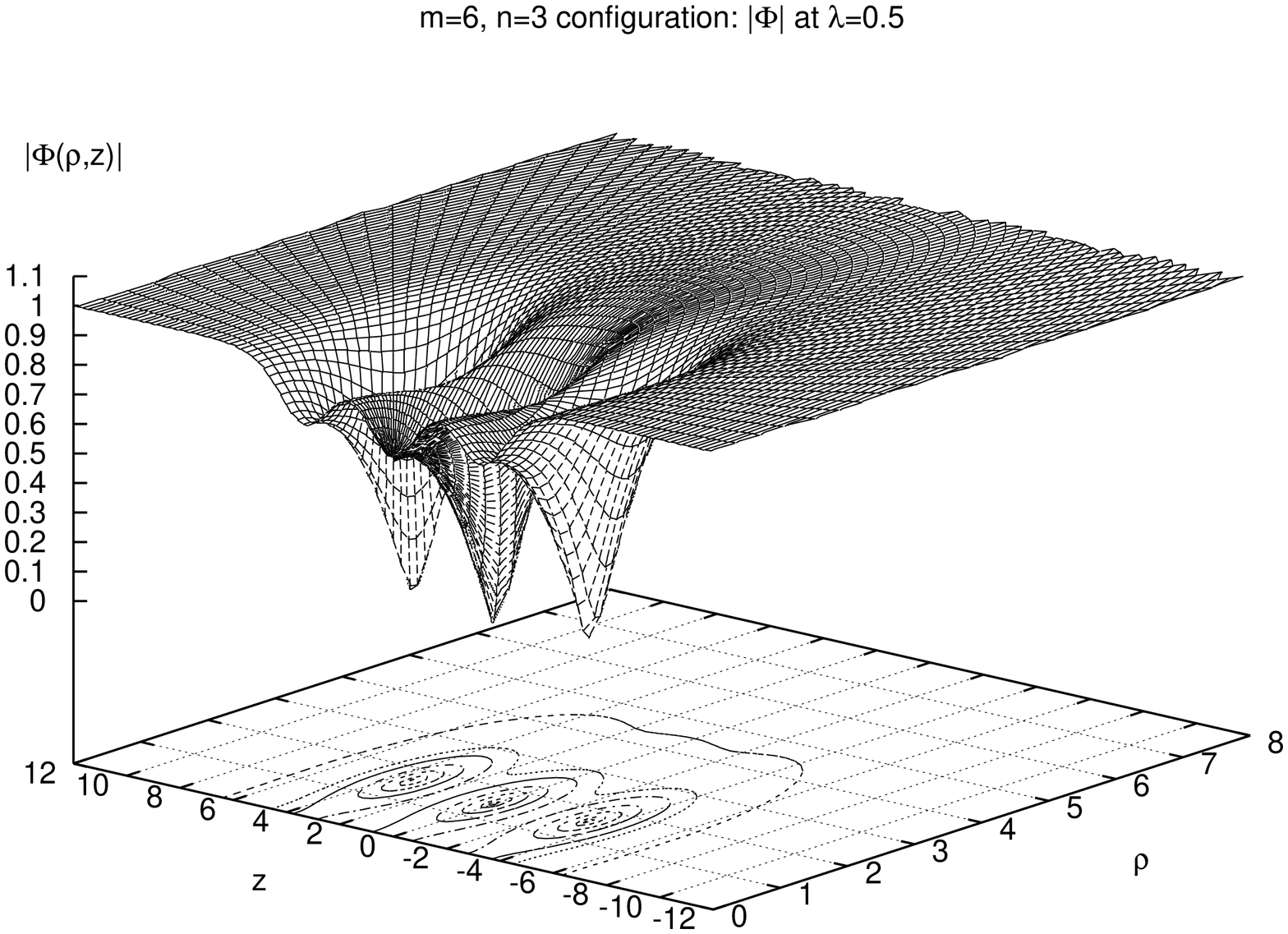,height=5.5cm, angle =-90}}}
\end{picture}
\caption{The rescaled energy density $E(\rho,z)$ and the modulus of the Higgs field 
$|\Phi(\rho,z)|$  are shown for the circular vortex 
solutions $m=2,n=3$ (single vortex), $m=4,n=3$ (double vortex) and  
$m=6,n=3$ (triple vortex) at $\lambda = 0.5$.
}
\end{center}
\end{figure}

The situation changes dramatically if $\vphi$ winding number  $n > 2$. 
The solution of the second type, which appears in that case 
are not multimonopole chains but sustems of vortex rings, in the BPS limit
either bounded 
with a single $n$-monopole placed at the origin, or without it \cite{KKS2}. 
One could expect that when charge of poles is increasing further beyond 
$n=2$ the chain solutions consisting of multimonopoles with winding number 
$n$ should exist, the monopoles and antimonopoles of the pairs
should approach each other further, settling at a still smaller equilibrium distance.

Constructing solutions with charge $n=3$ in the BPS limit ($\lambda = 0$)
however, we do not find chains at all.
Now there is no longer sufficient repulsion
to balance the strong attraction between 
the 3-monopoles and 3-antimonopoles.
Instead of chains, we now observe solutions with vortex rings,
where the Higgs field vanishes on closed rings around the symmetry axis.

Such a situation is not a novelty, for examply the rings of zeros of a
Higgs field are observed in Alice electrodynamics \cite{Bais}. It is also known that 
closed knotted vortices arise in theories with a non-trivial Hopf number \cite{Niemi}.
However the closed vortex solutions of Yang-Mills-Higgs theory were previously unknown.  

To better understand these findings let us consider
unphysical intermediate configurations, where we allow the
$\vphi$ winding number $n$ to continuously vary between the
physical integer values.
Beginning with the simplest such solution, the $m=2$ solution,
we observe, that the zeros of the solution with winding number $n$
continue to approach each other when $n$ is increased beyond 2,
until they merge at the origin. 
Here the pole and antipole do not annihilate, however. 
We conclude, that this is not allowed by the imposed symmetries and
boundary conditions.
Instead the Higgs zero changes its character completely,
when $n$ is further increased.
It turns into a ring with increasing radius for increasing $n$.
The physical 3-monopole-3-antimonopole solution in the BPS limit then
has a single ring of zeros of the Higgs field and no point zeros. The    
nodes of the Higgs field on the $xz$ plane have local indices $+1$, thus 
the $2d$ index computed on the spacial boundary in $i_\infty = 2$.

Considering the magnetic moment of the $m=2$ solutions,
we observe, that it is (roughly) proportional to $n$.
The pair of poles on the $z$-axis for $n=2$ clearly 
gives rise to the magnetic dipole moment of a physical dipole.
The ring of zeros also corresponds to a magnetic dipole field,
which however looks like the field of a ring of mathematical dipoles.
This corresponds to the simple picture that the positive and negative
charges have merged but not annihilated, and then spread out
on a ring. Thus, we can identify the $m=2$, $n=3$ solution with a closed 
vortex configuration. 

While the dipole moment of monopole-antimonopole chains with an equal number
of monopoles and antimonopoles has its origin in the magnetic charges of the 
configuration \cite{mapKK,KKS1}, the dipole moment of the closed vortices 
can be associated with loops of electric currents in analogy with the 
case of the sphaleron dipole moment \cite{Hindmarsh93}. 
Indeed, the electromagnetic current can be defined as 
$$
j_i^{em} = \partial_k F_{ik} \ ,
$$
where $F_{ik}$ is the `t Hooft field strength thensor (\ref{t-Hooft_F}) and there
are closed rings of currents inside of the core of the configuration \cite{KKS3}. 

Other solutions with even $\theta$ winding number
reside in the vacuum sector as well. 
For $m=2k>2$ solutions with zero scalar coupling it is now clear
how they evolve, when the $\vphi$ winding number is increased beyond $n=2$.
Starting from $k$ pairs of physical dipoles, 
the pairs merge and form $k$ vortex rings, which carry
the dipole strength of the solutions.

\subsubsection{$n>2$ and odd $m$: Vortices bounded with monopole}
The solutions with odd $\theta$ winding number have an isolated 
zero of the Higgs field at the origin, thus their
reside in the topological sector with charge $n$. 
For $m=4k+1$ the situation is somewhat similar to the above.
Here a single $n$-monopole remains at the origin,
whereas all other zeros form pairs, which for $n>2$ approach each other,
merge and form $2k$ rings carrying dipole strength. 
Since, however, a dipole on the positive axis and its respective
counterpart on the negative axis have opposite orientation, 
their contributions cancel in the total magnetic moment.
Thus the magnetic moment remains zero, as it must, because of the symmetry
of the ansatz \cite{KKS1,KKS3}. 
Non-zero scalar coupling does not change the situation but the picks of the 
energy density are getting sharper. 

\begin{figure}[t]
\begin{center}
\setlength{\unitlength}{1cm}
\lbfig{f-11}
\begin{picture}(0,3.0)
\put(-7.5,3.7)
{\mbox{
\psfig{figure=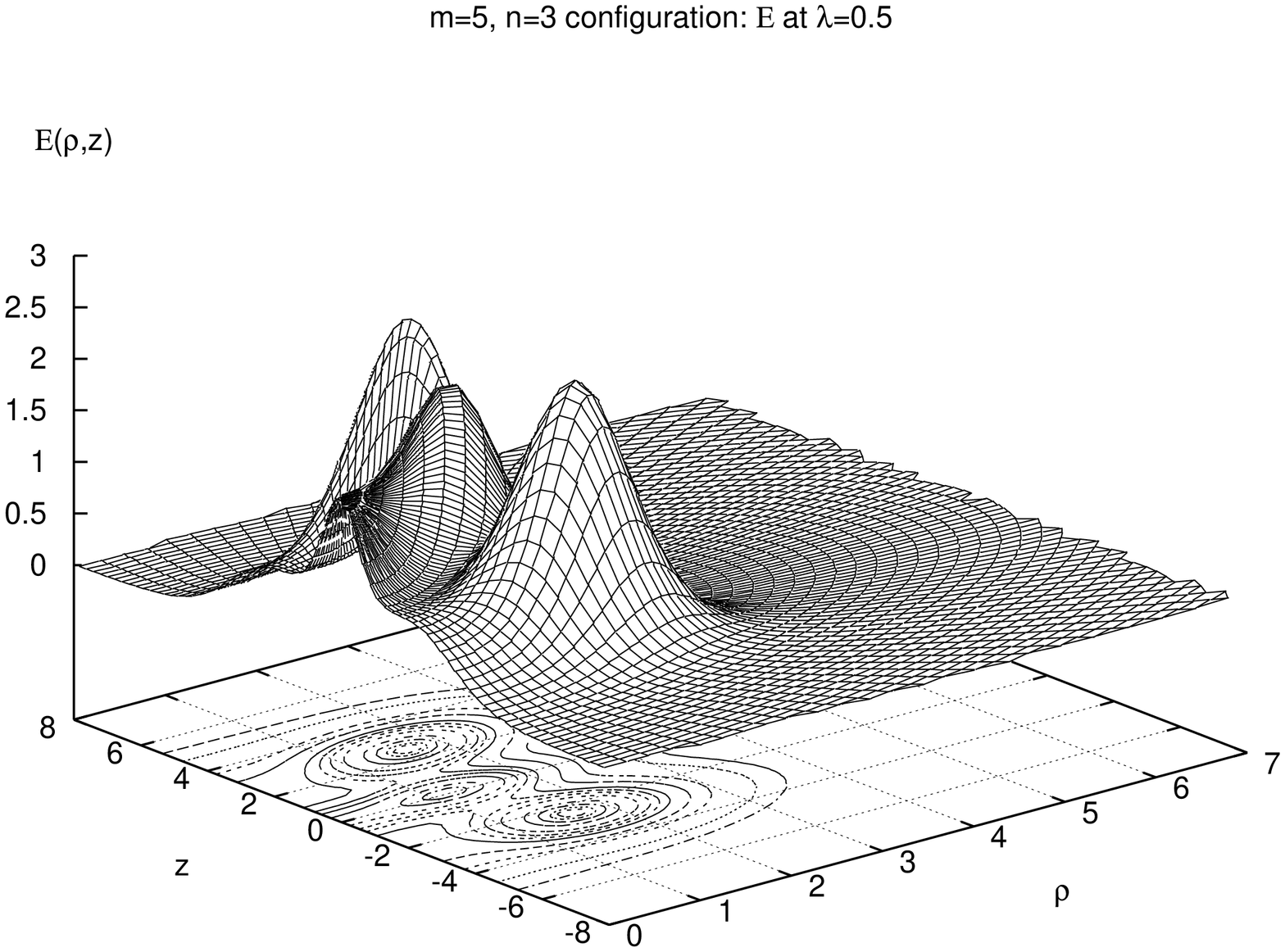,height=5.5cm, angle =-90}}}
\end{picture}
\begin{picture}(0,0.0)
\put(-2.5,3.7)
{\mbox{
\psfig{figure=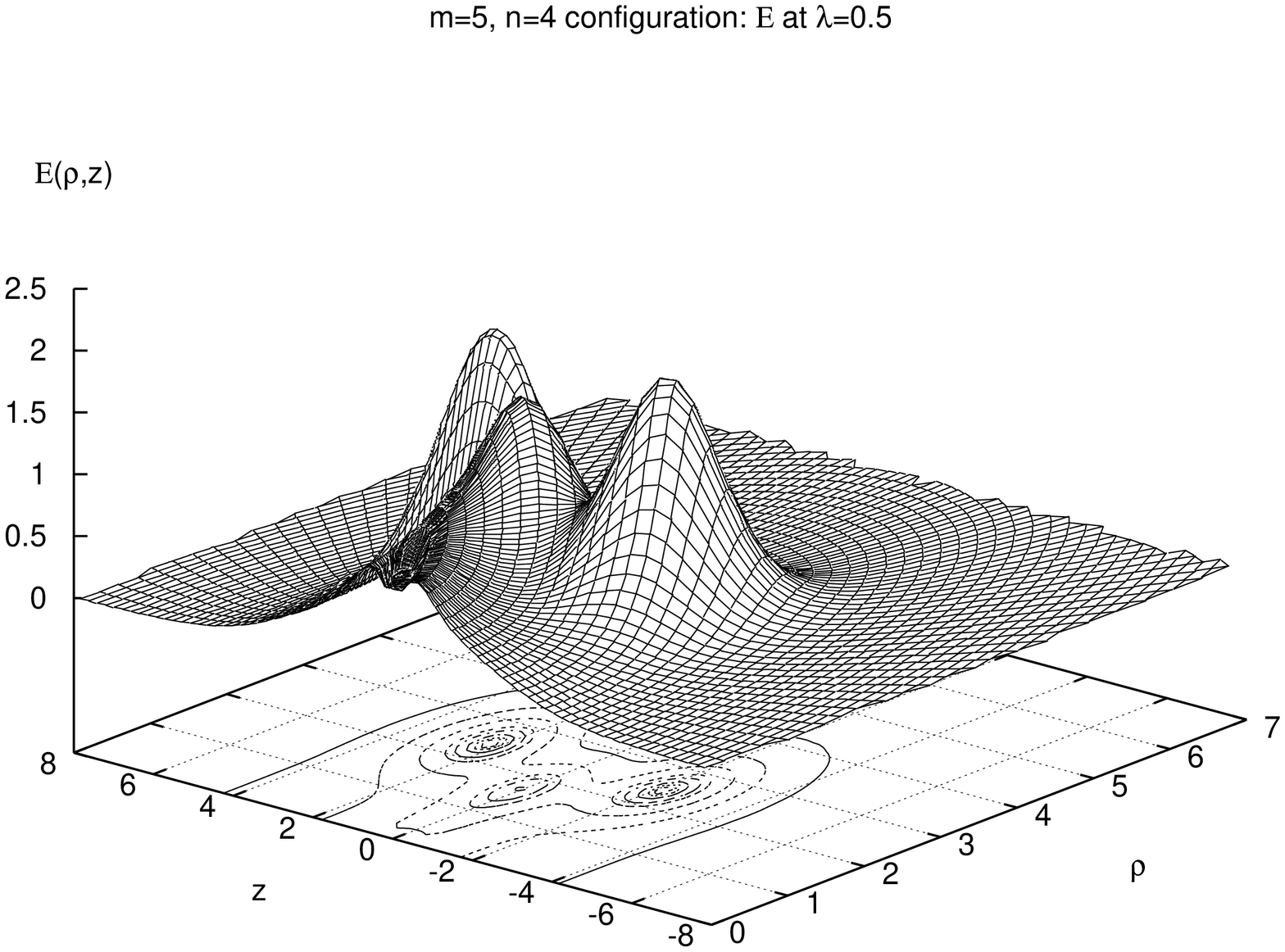,height=5.5cm, angle =-90}}}
\end{picture}
\begin{picture}(0,7.0)
\put(2.5,3.7)
{\mbox{
\psfig{figure=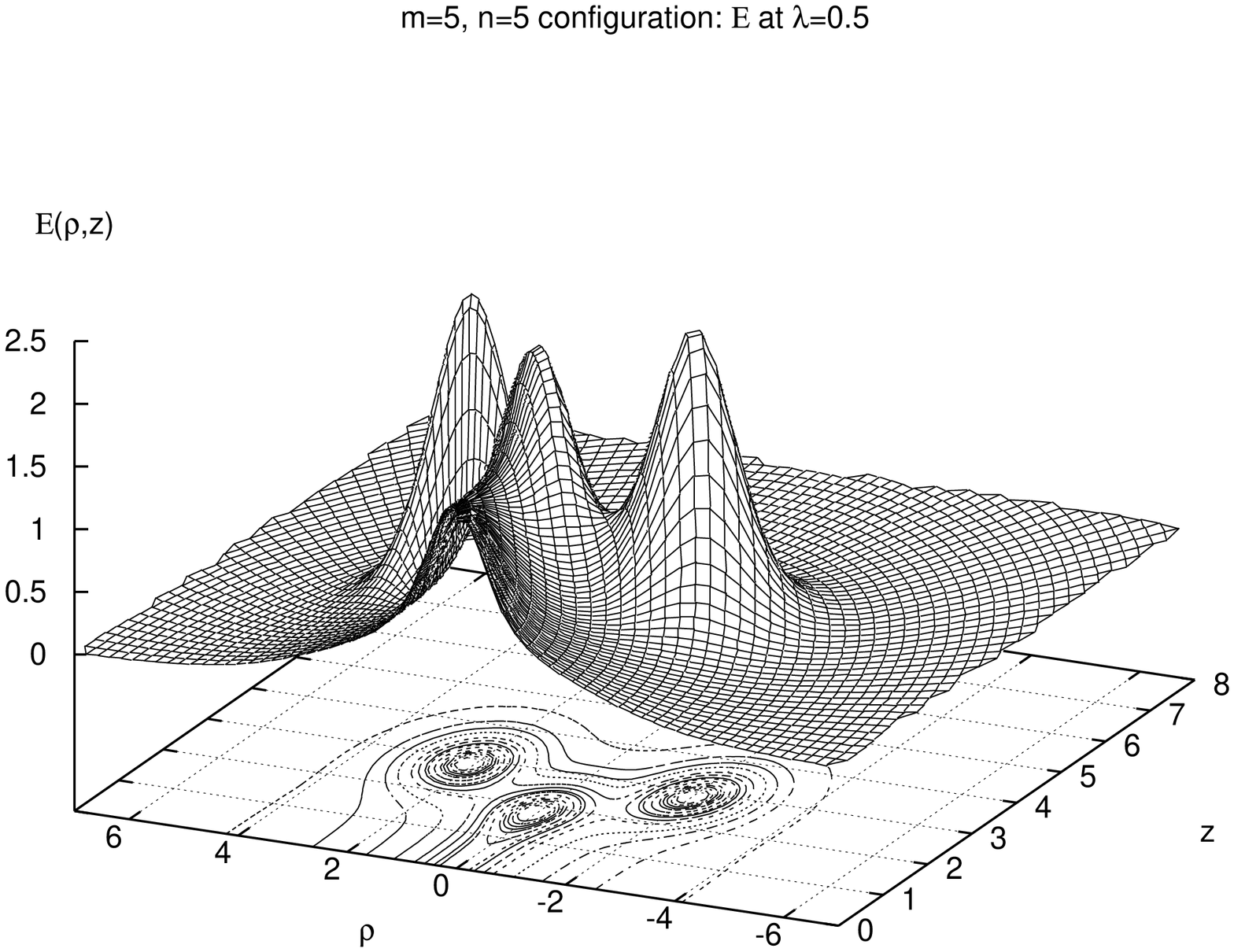,height=5.5cm, angle =-90}}}
\end{picture}
\begin{picture}(0,7.0)
\put(-7.7,7.5)
{\mbox{
\psfig{figure=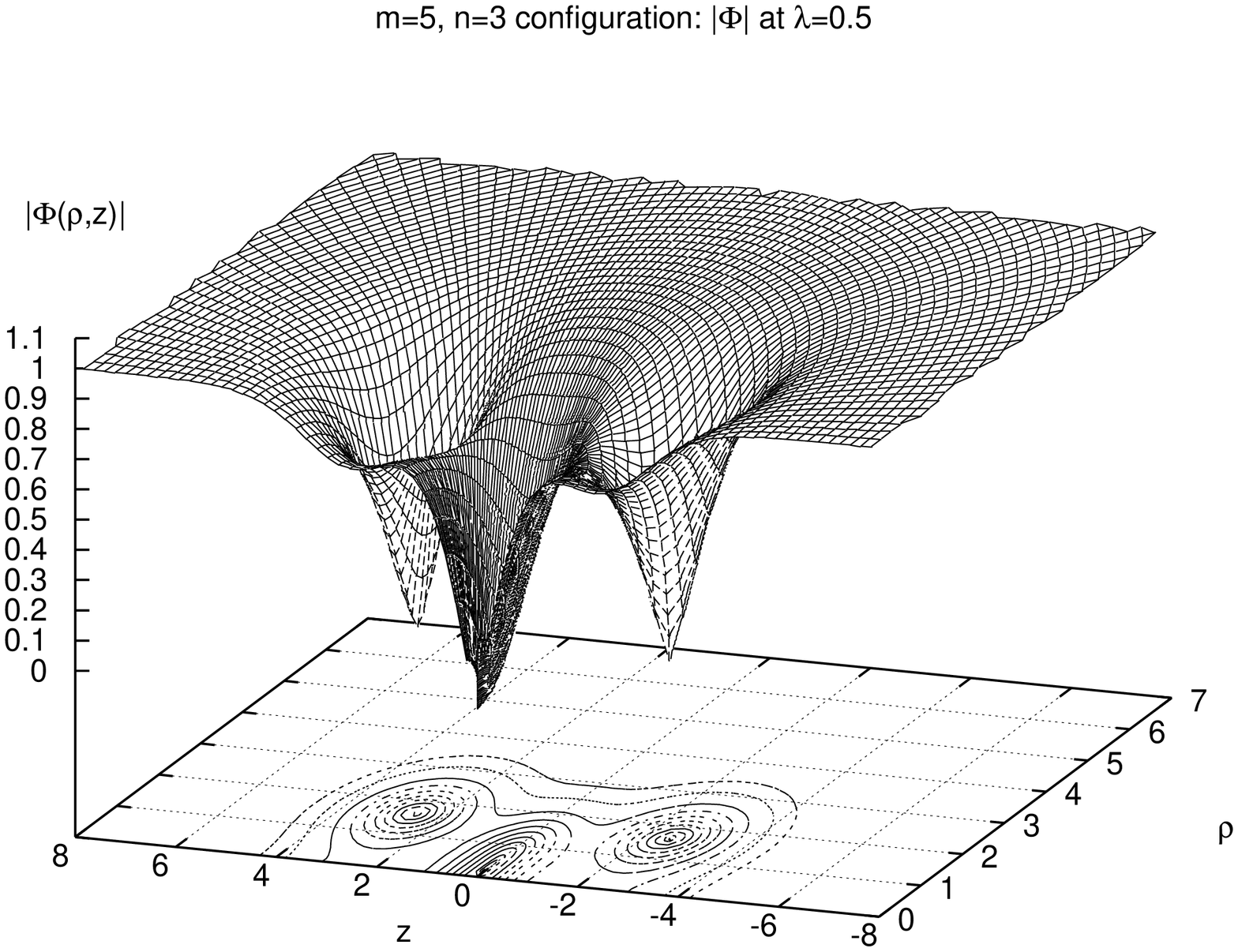,height=5.5cm, angle =-90}}}
\end{picture}
\begin{picture}(0,0.0)
\put(-2.7,7.5)
{\mbox{
\psfig{figure=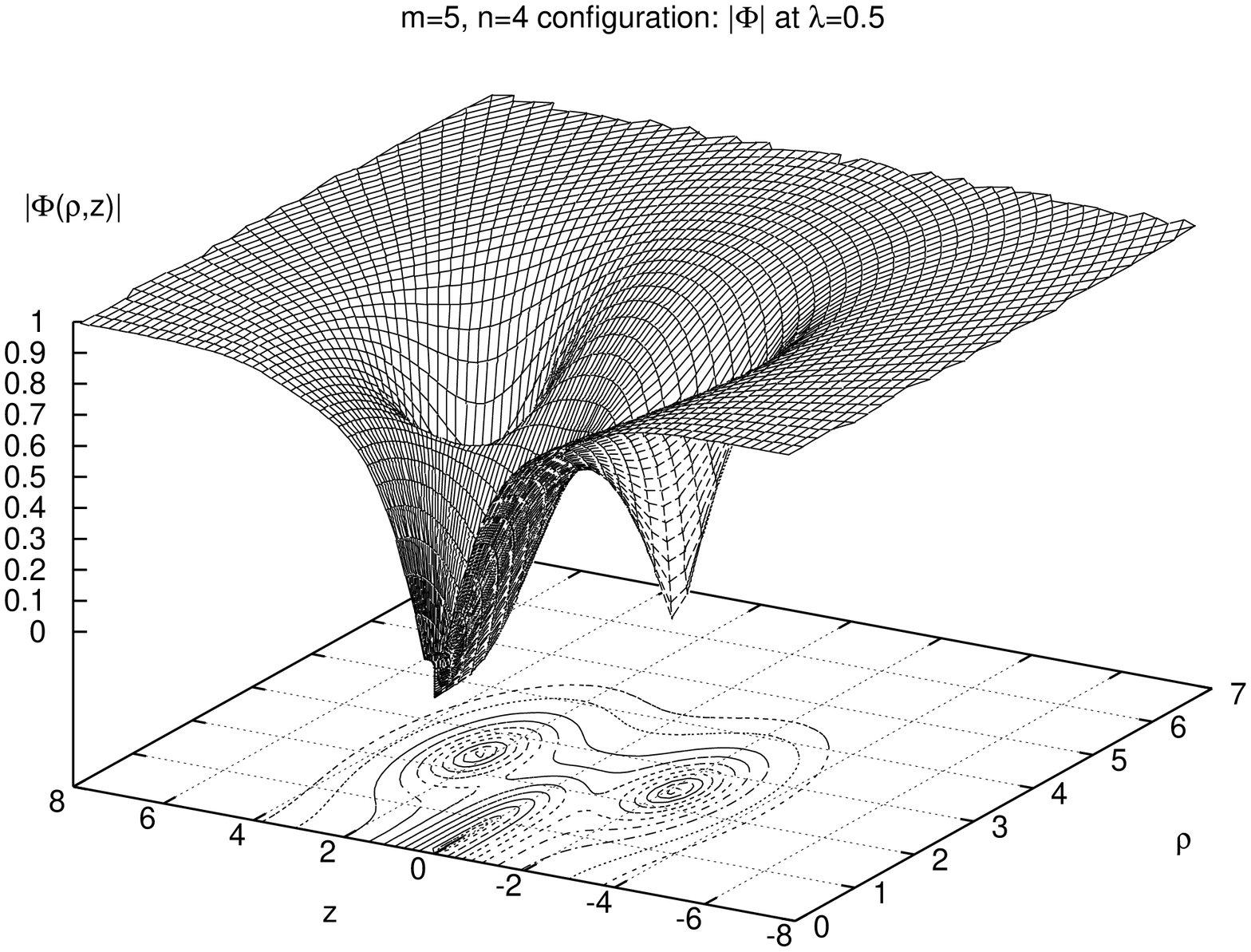,height=5.5cm, angle =-90}}}
\end{picture}
\begin{picture}(0,7.0)
\put(2.3,7.5)
{\mbox{
\psfig{figure=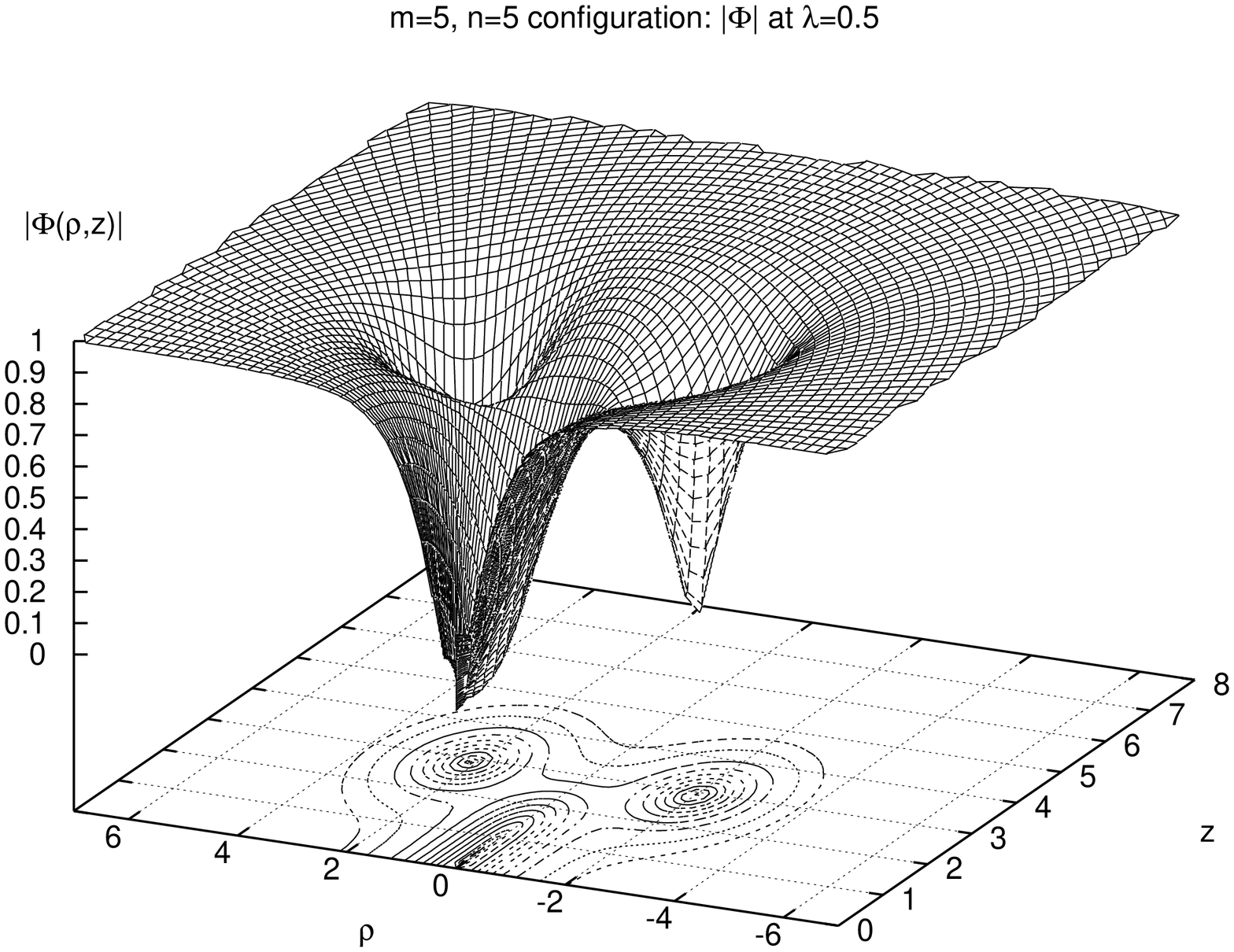,height=5.5cm, angle =-90}}}
\end{picture}
\caption{The rescaled energy density $E(\rho,z)$ and the modulus of the Higgs field 
$|\Phi(\rho,z)|$  are shown 
for the monopole-double vortex solution with winding numbers $m=5,n=3,4,5$ at $\lambda = 0.5$
}
\end{center}
\end{figure}

For $m=4k-1$, on the other hand, the situation is more complicated. Let us 
consider  the simplest case  $m=3$ in the limit $\lambda = 0$. Again, we consider unphysical configurations
with winding number $n$ continuously vary between $n=2$ chain solution  and $n=3$
configuration.  The difference from the case of an even $m$ is that 
in the initial state there are 3 poles on the $z$-axis,
which cannot form pairs, such that all zeros belong to a pair,
symmetrically located around the origin, thus the dipole moment of that configuration is 
zero. For $m=3$, we observe in the BPS limit, that
two vortices appear in the charge $n=3$ solution, emerging from
the upper and lower unpaired zero, respectively,
carrying opposite dipole strength. Thus, the local $2d$ index at the origin is 
$-1$ whereas other four indices 
are $1$, thus  $i_\infty = 3$. Increasing of scalar coupling decreases the radius of the vortices 
as well the distance between then. However both rings remain individual. 

This qualitative sketch of the properties of the new 
non-BPS axially symmetric multimonopoles and closed vortices is, of course, 
very superficial. We refer the reader to the papers  \cite{mapKK,KKS1,KKS2,KKS3}.
Nevertheless, there are many questions which remains to be studied. 
This work is currently underway.   

\paragraph{Acknowledgements}

I have always considered it a privilege to have had the opportunity of 
my study at the Department of Theoretical 
Physics, University of Belarus. The time I have spent there has been very fruitful.
I am very thankful to L.M.Tomilchik and E.A.Tolkachev, who were my teachers and 
advisors, for their valuable support, encouragement and guidance. 
They awaked my interest to the monopole problem.
I would like to thank I.D. Feranchuk for asking me to write this brief notes 
for this collection dedicated to the anniversary of my Alma Mater.  

This review is based on a work with Burkhard Kleihaus and Jutta Kunz \cite{KKS1,KKS2,KKS3}. 
I am ibdebted to Stephane Nonnemacher for helpful discussions, and in particular for 
clarifying the issue of the 2d index of vector fields. 
I am grateful to Ana Achucarro, Pierre van Baal,  Falk Bruckmann, 
Adriano Di Giacomo, Dieter Maison, and Valentine Zakharov   
for very useful discussions and remarks. 
I would like to acknowledge the hospitality 
at the Service de Physique Th\'eorique, CEA-Saclay and the Abdus Salam 
International Center for Theoretical Physics, where some parts of this work were 
carried out. I would like to thank RRZN in Cologne for computing time.

\end{document}